\documentclass[a4paper,11pt]{article}
\usepackage{jheppub} 
\usepackage{multirow}
\usepackage{rotating}
\usepackage{lineno}

\title{\boldmath A Universal Energy Relation between synchrotron and Synchrotron Self-Compton radiation in GRBs and Blazars}

\author[1,4]{Tao Wen}

\author[2,1,6,*]{Yu-Hua Yao}

\author[1,4]{Song-Zhan Chen}
\author[3]{Ben-Zhong Dai}
\author[1,4,5]{Yi-Qing Guo}

\affiliation[1]{Key Laboratory of Particle Astrophysics, Institute of High Energy Physics, Chinese Academy of Sciences, Beijing 100049, China}
\affiliation[2]{College of Physics, Chongqing University, No.55 Daxuecheng South Road, High-tech District, Chongqing, 401331, China}
\affiliation[3]{School of Physics and Astronomy, Yunnan University, Yunnan 650091, China}
\affiliation[4]{Tianfu Cosmic Ray Research Center, Chengdu 610213, Sichuan Province, China}
\affiliation[5]{College of Physics, University of Chinese Academy of Sciences, 100049, Beijing, China}
\affiliation[6]{Wisconsin IceCube Particle Astrophysics Center, University of Wisconsin–Madison, Madison, WI 53703, USA}

\emailAdd{yyao255@wisc.edu}

\abstract{The recent and brightest GRB 221009A observed by LHAASO marked the first detection of the onset of TeV afterglow, with a total of 7 GRBs exhibiting very high energy (VHE) afterglow radiation. However, consensus on VHE radiation of GRBs is still lacking. Multi-wavelength studies are currently a primary research method for investigating high-energy $\gamma$-ray astronomy. The limited sample of VHE GRBs, combined with their transient nature, hinders the progress of physical studies of GRBs. This paper aims to obtain useful information for GRB research through the properties of blazars, which share significant similarities with GRBs. By fitting high-quality and simultaneous multiwavelength spectral energy distributions with a one-zone leptonic model, the study explores the similarity of radiation properties of blazars and GRBs. A tight correlation between synchrotron and synchrotron self-Compton (SSC) emission luminosities suggests that blazars and GRBs share similar radiation mechanisms, to be specific, synchrotron radiation produces the observed X-ray photons, which also serve as targets for electrons in the SSC process. We hope that ground-based experiments can observe more GRBs in sub-TeV to confirm these findings.}

\begin{document}
\maketitle

\flushbottom

\section{Introduction} \label{sec:intro}
$\gamma$-ray bursts (GRBs) are the most energetic transient phenomena in the Universe. Despite being discovered decades ago \citep{1973ApJ...182L..85K}, many aspects related to their jet physics, particle acceleration and radiation processes, as well as radiation mechanisms, remain enigmatic (refer \citep{zhang2018physics} and therein). They are the main observational targets for various experimental facilities, both space-borne and ground-based. Recently, imaging atmospheric Cherenkov telescopes (IACTs) opened the very high energy (VHE, $>$100~GeV) window with the first TeV observation of GRB 190114C during the afterglow decaying phase \citep{2019Natur.575..455M}. Shortly thereafter, the Large High Altitude Air Shower Observatory (LHAASO), representing extensive air shower detectors \citep{2019arXiv190502773C}, successfully observed complete TeV radiation from the onset of GRB 221009A \citep{2023Sci...380.1390L}, which was the brightest GRB ever observed \citep{2023SciA....9J2778C}. These observations have brought unprecedented opportunities to study this still-mysterious class of transients in exquisite detail.

The current generation of $\gamma$-ray instruments is advancing our understanding of GRBs as high-energy sources, particularly through multiwavelength joint observations \citep{2022Galax..10...66M,2022Galax..10...67B}. However, the transient nature of GRBs on the timescale of seconds presents challenges in obtaining simultaneous or at least contemporaneous data across different wavelengths. Among the 7 GRBs reported with VHE emission \footnote{http://tevcat.uchicago.edu/}, over half of them either lacked available VHE data or simultaneous other-wavelength SED data, making it difficult to reach a consensus on the radiation mechanism. Multifrequency observations of GRB 190114C support the possibility that inverse Compton emission is commonly produced in GRBs \citep{2019Natur.575..459M}. Additionally, GRB 190829A requires complicated jet or radiation configurations for interpretation \citep{2021Sci...372.1081H}. Moreover, besides the common radiation mechanisms (leptonic or hadronic emission), the detection of $\gamma$-rays up to 13 TeV from GRB 221009A even triggered models of many exotic origins \citep{2022JETPL.116..767T,2023PhRvL.131y1001G,2023PhRvD.107h3038B,2023PhRvD.108b3002W,2023ChPhL..40a1401Z}.

Considering the sparse VHE GRB samples, it is likely to study VHE emission and related radiation mechanisms and jet properties by leveraging similar sources, such as AGNs. Both GRBs and AGNs are generated in relativistic jets powered by their central black holes. GRBs are associated with ultra-relativistic jets launched during the core collapse of massive stars or mergers of two compact objects, while AGN are believed to be fueled by super-massive rotating black holes that launch mildly relativistic jets, with those pointing towards the observer being classified as blazars.

In contrast to GRBs, there have been numerous observations of $\gamma$-ray blazars, most of which have been extensively researched. Modeling the broad-band spectral energy distribution (SED) of a source is a common method used to investigate source physics. It is widely recognized that the SED of all known $\gamma$-ray blazars exhibits double-humped structures, suggesting the involvement of at least two distinct physical emission processes. The first peak, spanning from radio to soft X-ray frequencies, is likely attributed to synchrotron emission from relativistic electrons in the jet, while the second component is commonly believed to be Synchrotron Self-Compton (SSC) scattering emission from the same electron population \citep{1992ApJ...397L...5M,1996A&AS..120C.503G}. However, alternative models such as external Compton models \citep{1992A&A...256L..27D}, and more intricate scenarios involving hadronic processes, have also been put forward as plausible explanations \citep{1993A&A...269...67M}.

Many studies have delved into and demonstrated the similarities between GRBs and blazars, examining various aspects such as jet power, synchrotron luminosity, radiation efficiency, and Doppler factor, suggesting that they may share similar jet physics \citep{2011ApJ...740L..21W,2012Sci...338.1445N,2014ApJ...793...36L,2014ApJ...786L...8W,2013ApJ...774L...5Z}. Radiation characteristics have also been investigated through analogies, focusing on the peak or energy of synchrotron radiation and its correlation with the physical properties of the jet \citep{2013ApJ...774L...5Z,2014ApJ...780L..14M,2019ApJ...873..120Z}, or studying the correlation between X-rays and optical or radio emissions under synchrotron radiation \citep{2022MNRAS.509.4143Z}.


In this paper, we utilize analogies to compare the radiation power of X-rays and VHE energy bands of blazar flares and GRBs,  in order to investigate the mechanisms behind VHE radiation in GRBs, as reaching a definitive conclusion about their VHE emission has been challenging up to now. The paper is organized as follows: Section \ref{sec:data} offers a succinct description of the blazar and GRB samples. In Section \ref{sec:method}, we delineate the SED modeling within the framework of a homogeneous, one-zone SSC mechanism, and derive the best-fitting parameters using the MCMC package. Subsequently, Section \ref{sec:results} presents the findings, with a summary provided in Section \ref{sec:summary}.

\section{Data sample}\label{sec:data}
In this section we briefly introduce the data sample we used, two kinds of sample were utilized and listed in Table \ref{tab:candi} in this study: blazars and GRBs.

\begin{sidewaystable*}[!htp]
\footnotesize
\centering
\caption{List of Sources and References}
\vspace{0.2cm}
\begin{tabular}{cccccc}
\hline
Source                   & Redshift                  & Data Time              & State & Data and Refs & EBL correct\footnote{"obs" represents the observed spectrum data. If the observed data is de-absorbed using the EBL model, the corresponding EBL model will be listed here.} \\
\hline
\multirow{4}{*}{Mrk 421} & \multirow{4}{*}{0.031} & 55242-55245 & flaring  & Swift-UVOT, Swift-XRT, Swift-BAT, Fermi-LAT, ARGO-YBJ \citep{2016ApJS..222....6B} & obs \\
                         &                           &   55265    & flaring      &  SMA,Fermi-LAT,MAGIC,VERITAS,Swift-XRT,RETE,UVOT,(B,V,R,I) \citep{2015AA...578A..22A}  & obs \\
                         &                        &   56130-56187  & flaring & MAXI-GSC,  Swift-BAT, Fermi-LAT, ARGO-YBJ \citep{2016ApJS..222....6B} & obs \\
                         &                           &   57757    & flaring      &    Swift-UVOT, Swift-XRT, Swift-BAT, NuSTAR, Fermi-LAT, MAGIC, FACT \citep{2021AA...655A..89M}  & obs \\
\hline
\multirow{4}{*}{Mrk 501} & \multirow{4}{*}{0.034}    &   54911-54918    & outburst      & Suzaku, Fermi-LAT, VERITAS, MAGIC\citep{2011ApJ...729....2A}   & obs \\
                         &                          &   54952-54955 &   flaring    &  Whipple, VERITAS, Fermi-LAT, Swift-XRT \citep{2016AA...594A..76A}   & obs \\
                         &                           &   56858    & flaring      &  MAGIC, Fermi-LAT, Swift-XRT, Swift-BAT\citep{2020AA...637A..86M}   & obs \\
                         &                           &   56865   & flaring      &   MAGIC, Fermi-LAT, Swift-XRT, Swift-BAT \citep{2020AA...637A..86M}   & obs \\
\hline
   1ES 1959+650          &               0.048       &  53879-53881    & flaring(only x)    &  Optical-UV, Suzaku, Swift-XRT, MAGIC \citep{2008ApJ...679.1029T}  & obs \\
\hline
   PKS 2155+304          &               0.116       &  53945    & flaring            &   Optical-V, Chandra, RossiXTE, H.E.S.S. \citep{2009AA...502..749A}  & \citep{2008AA...487..837F} \\
\hline
      W Comae              &              0.102        &  54624-54626    & flaring    & Swift-UVOT, Swift-XRT, VERITAS \citep{2009ApJ...707..612A}  & obs \\
\hline
   3C 66A                &            0.444          &  54743-54745    & flaring    &  MDM, UVOT, Swift-XRT,  Fermi-LAT, VERITAS \citep{2011ApJ...726...43A} & \citep{2008AA...487..837F} \\
\hline
     1ES 1727+502        &                 0.055     &  57090–57440    & flaring    & Fermi-LAT, MAGIC, Swift-XRT, Swift-UVOT \citep{2022MNRAS.515.2633P}  & \citep{2008AA...487..837F} \\
\hline
   GRB 190114C       &                 0.4245    &  $T_0$+68 s-$T_0$+110 s     & afterglow  & Swift-XRT, Swift-BAT, Fermi-LAT, Fermi-GBM, MAGIC\citep{2019Natur.575..455M} & obs \\
\hline
    GRB 190829A          &                  0.0785   &  $T_0$+4.3 hrs-$T_0$+7.9 hrs  &    afterglow  &Swift-XRT, H.E.S.S. \citep{2021Sci...372.1081H} & obs \\
\hline
   GRB 221009A           &                0.151      &  $T_0$+900 s-$T_0$+ 2000 s     &  afterglow & HXMT, LHAASO \citep{2023Sci...380.1390L,2023arXiv230301203A} & obs \\
\hline
\hline
    \end{tabular}
    \label{tab:candi}
\end{sidewaystable*}

\subsection{blazars}
There are a total of 7 VHE $\gamma$-ray blazars listed, making them ideal candidates for detailed studies that cannot be carried out for other blazars that are fainter, located farther away, or have complicated structures. Mrk 421, located at \textit{z}=0.031, is one of the brightest known VHE blazars. It was the first extragalactic source detected by IACTs. In addition to the abundant observations provided by IACTs such as MAGIC and VERITAS, extensive air shower observatory ARGO-YBJ continuously monitored it for five years, reporting seven flaring episodes, among which we selected the largest flare and another one for analysis. Mrk 501, located at z=0.034, is another well-known VHE blazar regularly monitored by a wide range of multi-wavelength instruments. It has VHE $\gamma$-ray observations from instruments such as Whipple, VERITAS, and MAGIC, and we also selected its 4 flaring states for SED modeling.

W Comae had a strong outburst of VHE $\gamma$-ray emission in 2008 June and was detected by VERITAS \citep{2009ApJ...707..612A}, which detected 3C 66A in a flaring state in October of the same year \citep{2011ApJ...726...43A}. Both sources were carried out with multi-wavelength observations in comparable timescales. Simultaneous flaring was also observed from TeV blazar 1ES 1727+502 and 1ES 1959+650 in the X-ray and $\gamma$-ray bands \citep{2022MNRAS.515.2633P,2008ApJ...679.1029T}. Here is a interesting for the 1ES 1959+650,  since the source exhibited a relatively high state in X-rays and optical, while in the
VHE band it was at one of the lowest level \citep{2008ApJ...679.1029T}. PKS 215+304 is among the brightest and most extensively studied BL Lac objects in the southern hemisphere across various wavelengths. It has been detectable by HESS almost nightly since 2002 \citep{2005A&A...430..865A,2009AA...502..749A}.

\subsection{GRBs}
On the TeVcat website \footnote{http://tevcat.uchicago.edu/}, there are 7 VHE GRBs listed. However, with the exception of GRB 190114C, GRB 190829A, and GRB 221009A, the other VHE-announced GRBs either lacked available VHE data or simultaneous other-wavelength SED data. GRB 190114C was the first GRB reported with simultaneous observations up to the TeV range by IACT. The VHE afterglow phase of GRB 190829A was detected between 4 and 56 hours after the onset of prompt emission, despite its low luminosity and redshift \citep{2021Sci...372.1081H}. It is also one of the nearest GRBs observed ever. Meanwhile, GRB 221009A has been described as the brightest ever detected \citep{2023Sci...380.1390L}, with $\gamma$-rays observed up to 13 TeV by the LHAASO experiment \citep{2023SciA....9J2778C}.

\section{Model and method}
In this section, we will first provide a brief introduction to SED modeling. This will be followed by a description of the procedures adopted for obtaining the best fitting parameters and determining the luminosity in the X-ray and VHE range.
\subsection{SED modeling: one-zone SSC model}\label{sec:method}
We assume that an emitting zone, which is a spherical blob, moves downstream of the jet with a bulk Lorentz factor $\Gamma_b$. This blob is filled homogeneously with electrons in a randomly oriented magnetic field with a mean magnetic field strength of $B^{\prime}$. We also assume that the jet axis is at an angle $\Theta=1/\Gamma_b$, and for simplicity, we take the Doppler factor $\delta$ to be equal to $\Gamma_b$. Quantities in the observer’s frame are unprimed, and quantities in the frame comoving with the jet blob are primed.

Ultra-relativistic electrons are assumed to be instantaneously accelerated into a power-law distribution in electron energy
\begin{equation}\label{eq:powerlaw}
    N_{e}(\gamma)=K_e\,\gamma^{-p1}.
\end{equation}
The combination of radiative cooling and particle escape from the emission region leads to a broken power-law equilibrium distribution, where the electrons follow:
\begin{equation}\label{eq:broken powerlaw}
    N_{e}(\gamma)=
    \left\{
    \begin{aligned}
        & K_e\,(\gamma/\gamma_{break})^{-p1}, \quad \gamma < \gamma_{break} \\
        & K_e\,(\gamma/\gamma_{break})^{-p2}, \quad \gamma > \gamma_{break} \\
    \end{aligned}
    \right.
\end{equation}
with indices $p_1$ and $p_2$ below and above the break $\gamma_{break}$, within the low- and high-energy cut-offs, denoted as $\gamma_{min}$ and $\gamma_{max}$.

The synchrotron radiation flux of relativistic electrons can be expressed as \citep{1986A&A...164L..16C}:
\begin{equation}
    f_{\epsilon}^{\mathrm{syn}}=\frac{\delta_{\mathrm{D}}^{4}\epsilon'J_{\mathrm{syn}}^{\prime}(\epsilon')}{4\pi d_{L}^{2}}=\frac{\sqrt{3}\delta_{\mathrm{D}}^{4}\epsilon'e^{3}B^{\prime}}{4\pi h d_{L}^{2}}\int_{1}^{\infty}d\gamma'N_{e}^{\prime}(\gamma')R(x)
\end{equation}
where \textit{$\delta_{\mathrm{D}}$} is the Doppler factor, \textit{$\epsilon^{\prime}$} represents the dimensionless energy of the emitted photon, \textit{$\gamma^{\prime}$} is the electron's Lorentz factor, \textit{e} denotes the fundamental charge, $\textit{h}$ is Planck's constant, and $d_L$ signifies the luminosity distance given by
\begin{equation}
    d_L=\frac{(1+z)c}{H_0} \int_0^{z}\frac{dz^{\prime}}{\sqrt{\Omega_M(1+z^{\prime})^3+\Omega_\lambda}}
\end{equation}
with $\Omega_M$=0.3, $\Omega_\lambda$=0.7, $H_0$ =$\rm 70~km~s^{-1}~Mpc^{-1}$ \citep{2022PTEP.2022h3C01W}.

The function $R(x)$ is defined as:
\begin{equation*}
     R(x) = \frac{{x}}{{2}} \int_{0}^{\pi} d\theta \sin(\theta) \int_{\frac{{x}}{{\sin\theta}}}^{\infty} dt K_{\frac{{5}}{{3}}}(t)
\end{equation*}
where $x = \frac{{4\pi\epsilon'm_e^2c^3}}{{3eB^{\prime}h\gamma'^2}} $,  $\theta$ is the pitch angle between the magnetic field and the motion of the electron, and $K_{5/3}$ represents the modified Bessel function of the second kind of order 5/3.

The observed SSC flux is
\begin{equation}
\begin{aligned}
f_\epsilon^{\mathrm{SSC}}= & \frac{3}{4} c \sigma_{\mathrm{T}} \epsilon_s^{\prime 2} \frac{\delta_{\mathrm{D}}^4}{4 \pi d_L^2} \int_0^{\infty} d \epsilon^{\prime} \frac{u^{\prime}\left(\epsilon^{\prime}\right)}{\epsilon^{\prime 2}} \\
& \times \int_{\gamma_{\min }^{\prime}}^{\gamma_{\max }^{\prime}} d \gamma^{\prime} \frac{N_e^{\prime}\left(\gamma^{\prime}\right)}{\gamma^{\prime 2}} F_{\mathrm{C}}\left(q^{\prime}, \Gamma_e^{\prime}\right),
\end{aligned}
\end{equation}
with $\epsilon_s^{\prime}$ is the scattered photon's dimensionless energy and $u^{\prime}$ is the radiation energy density \citep{2008ApJ...686..181F}. Here we consider the full Compton scatter cross section (i.e. the Klein-Nishina cross section is taken into account), $F_{\mathrm{C}}$ is the Compton scattering kernel for isotropic photon and electron distributions \citep{1968PhRv..167.1159J,1970RvMP...42..237B}
\begin{equation*}\label{eq:fc}
\begin{aligned}
F_{\mathrm{C}}\left(q^{\prime}, \Gamma_e^{\prime}\right)= & {\left[2 q^{\prime} \ln q^{\prime}+\left(1+2 q^{\prime}\right)\left(1-q^{\prime}\right)\right.} \\
& \left.+\frac{1}{2} \frac{\left(\Gamma_e^{\prime} q^{\prime}\right)^2}{\left(1+\Gamma_e^{\prime} q^{\prime}\right)}\left(1-q^{\prime}\right)\right] H\left(q^{\prime} ; \frac{1}{4 \gamma^{\prime 2}}, 1\right).
\end{aligned}
\end{equation*}
with  $q'\equiv\frac{\epsilon_s'/\gamma'}{\Gamma_e'(1-\epsilon_s'/\gamma')}$, $\Gamma_e'=4\epsilon'\gamma'$, and $H(x; a, b)$ is the Heaviside function defined as $H = 1$ if $a \leq x \leq b$ and $H = 0$ otherwise.

In short, there is a set of specific parameters (magnetic filed $B^{\prime}$, Doppler facor $\delta$, electron spectral index $p1$, $p2$, $\gamma_{break}$, normalization $K_e$, and electron Lorentz factor cut-off $\gamma_{max}$) for the SED modeling of one source and need to be determined experimentally by fitting the data.

\subsection{Fitting strategy and procedure}
The maximum likelihood method is used to obtain the optimal values of free parameters. In general, data measurements and errors follow a Gaussian distribution, which allows us to construct a likelihood function:
\begin{equation}\label{eq:sed ts}
	L(\theta|f_1,f_2,\cdots,f_n) = \prod_{i=1}^{N} \frac{1}{\sqrt{2\pi\sigma_i^2}}\exp\left[-\frac{(f_i-f_{i}^{'}(\theta))^2}{2\sigma_i^2}\right]
\end{equation}
where $f_1,f_2,\cdots,f_n$ is the observed data, N is the number of observed data points, $\sigma_i$ represents the error of data, $f_i'$ denotes the model value with the parameters $\theta$(\textit{B}, $\delta$, $p_1$, $p_2$, $\gamma_{break}$, $\gamma_{max}$ and $K_{e}$).
Technically, the Markov Chain Monte Carlo (MCMC) technique is used to establish the parameter and its error, and the public Python package Emcee \citep{2013ascl.soft03002F} is adopted for fitting the best parameters.
The extragalactic background light (EBL) absorption of the VHE $\gamma$-ray is take account into calculation, according to the EBL model \citep{2010ApJ...712..238F}, which is favored by LHAASO for the observation of GRB 221009A. If the flux points are de-absorbed using the EBL model, we will first correct them using the corresponding EBL model, and then proceed to fit the SSC model uniformly using the EBL model.

{

To calculate the energy released by the synchrotron and SSC radiation processes, we utilize the following equation \citep{2010ApJ...711.1073U}
\begin{equation}
     L_{iso} = 4~\pi~d_L^2 f_{obs}
\label{eq:L}
\end{equation}
where $f_{obs}$ denotes the observed peak flux for energy range in the source-frame, which is calculated as:
\begin{equation}
    f_{obs}=\int_{E_1/(1+z)}^{E_2/(z+1)}N(E)EdE
\end{equation}
For each source, the peak values of the two peaks are initially identified. Given that observational data typically cluster within 1.5 orders of magnitude before and after the peaks, this range is adopted as the upper and lower limits for energy integration, and the flux is integrated over this energy range. It is important to note that beyond this energy range, the flux sharply decreases, therefore, hence extending the energy range would not significantly impact the integral results.

\section{Result}\label{sec:results}

\begin{figure*}[!htp]
    \centering
    \includegraphics[width=0.95\textwidth]{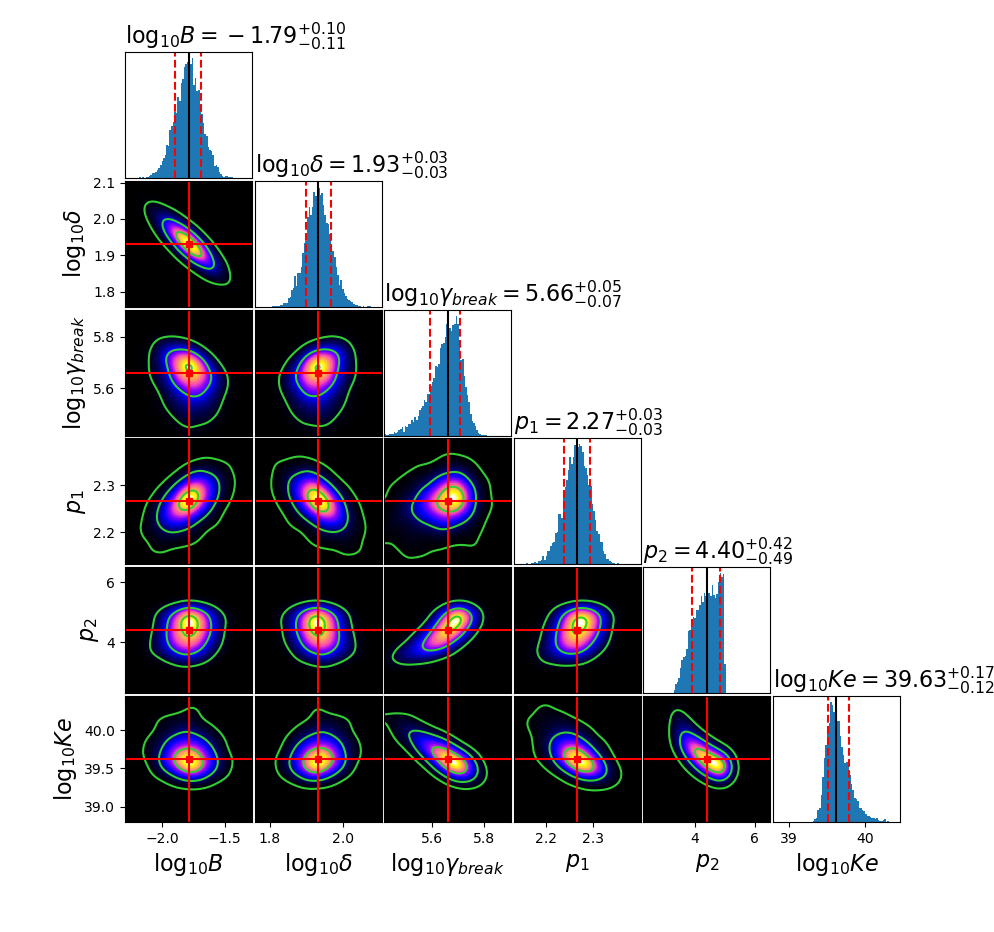}
    \caption{Posterior distribution contours of the one-zone SSC model parameters derived from our MCMC fits for Mrk 421 flaring event that occurred at \textit{MJD}=55265. }
    \label{fig:mcmc}
\end{figure*}

\begin{figure*}[!htp]
    \centering
    \includegraphics[width=0.244\textwidth]{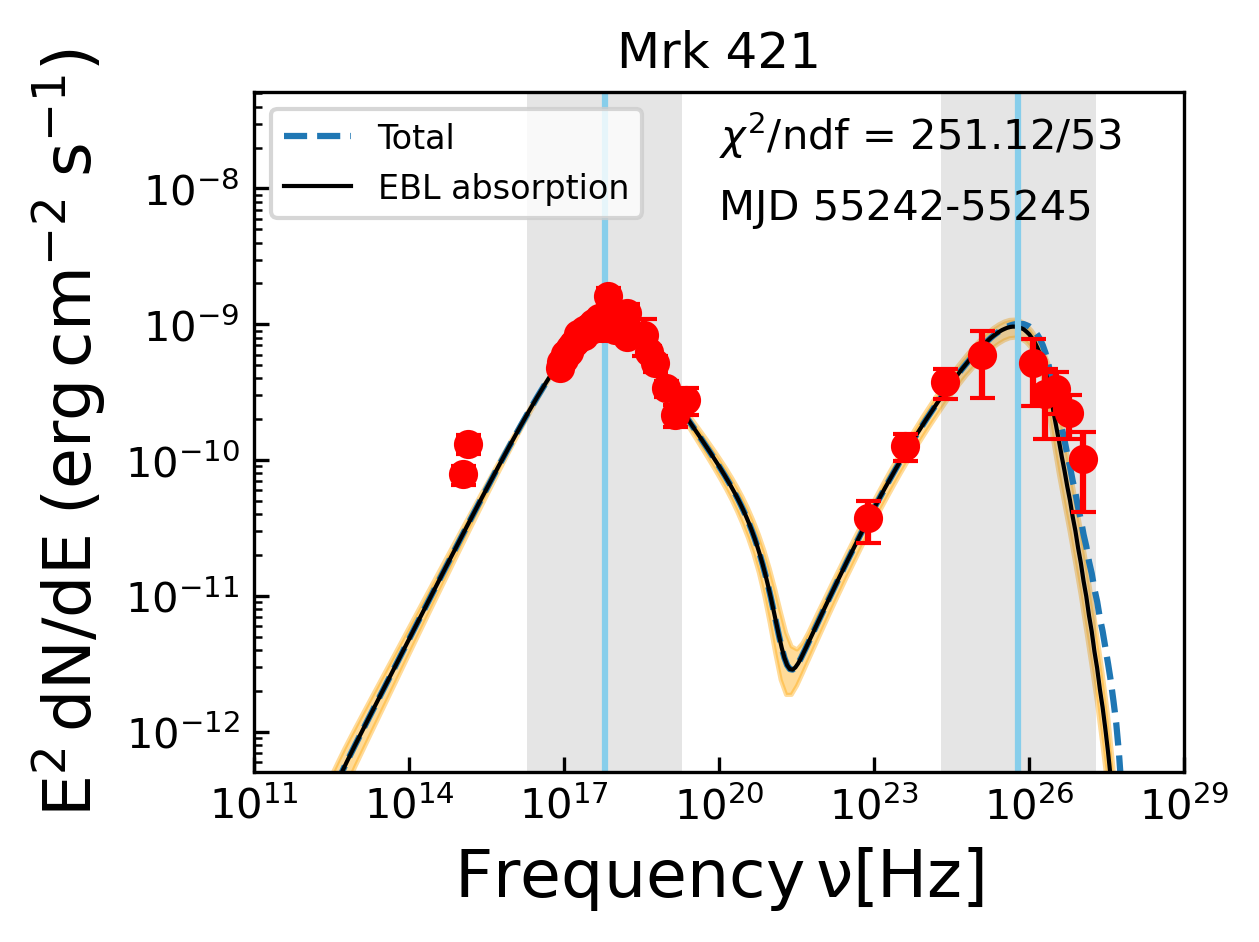}
    \includegraphics[width=0.244\textwidth]{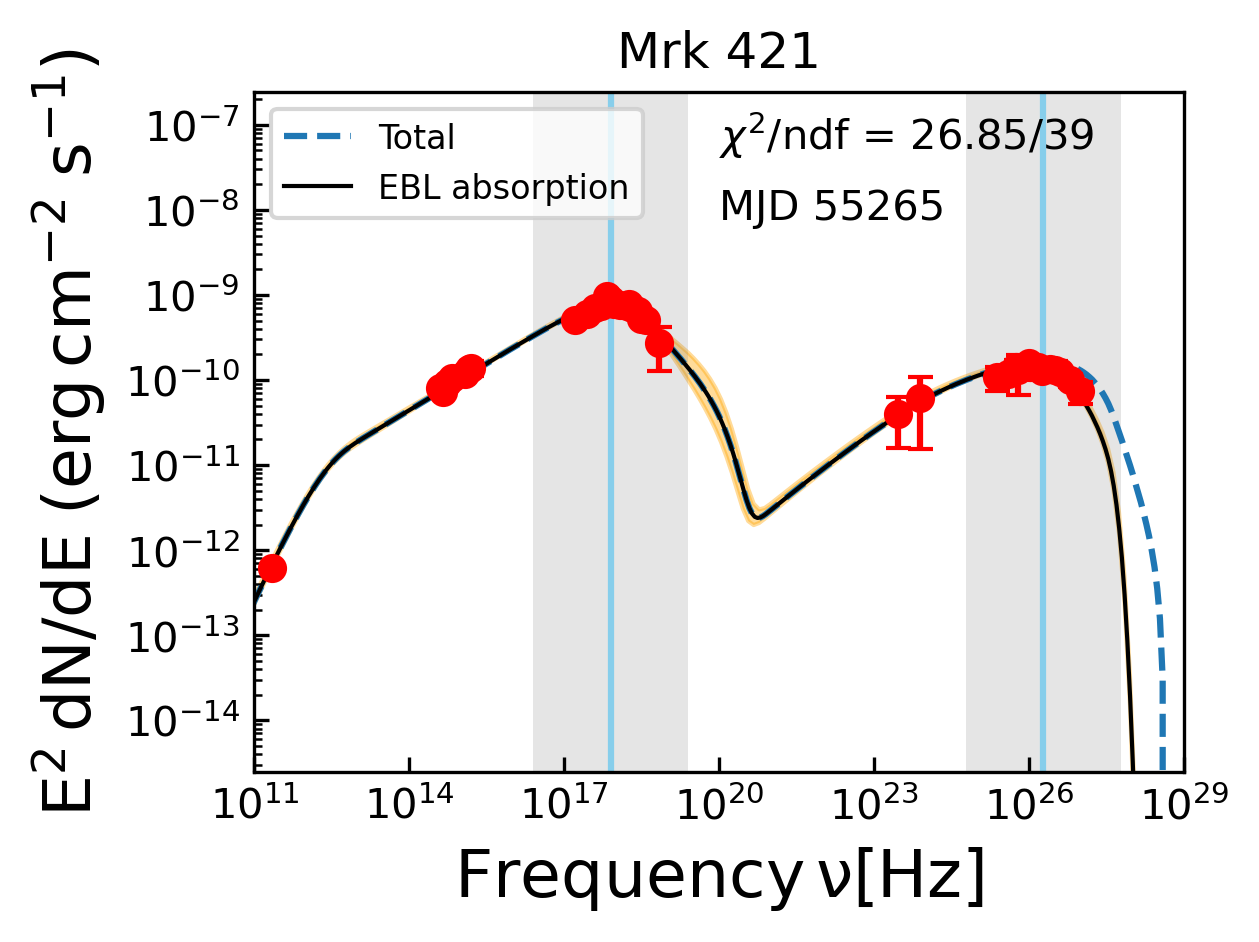}
    \includegraphics[width=0.244\textwidth]{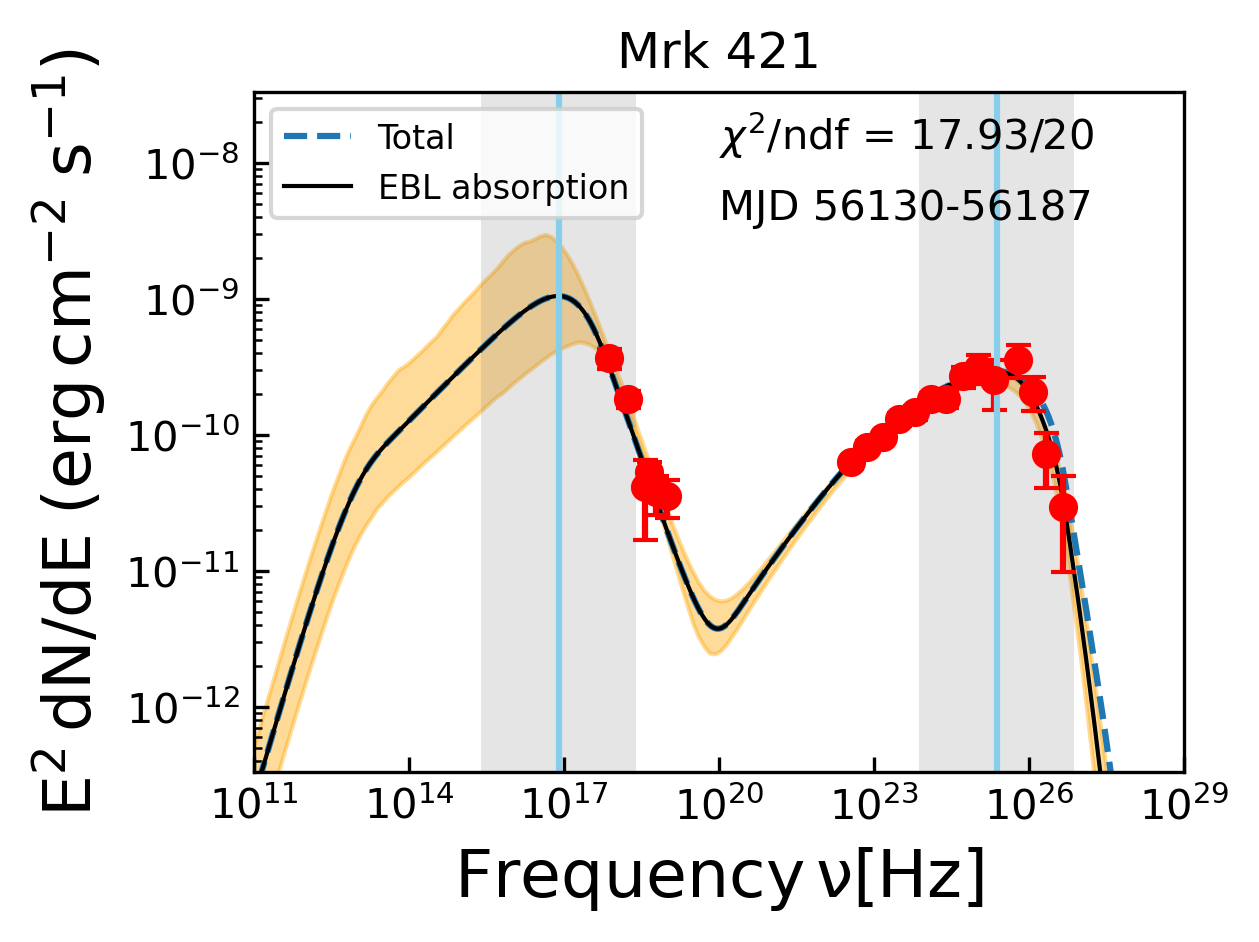}
    \includegraphics[width=0.244\textwidth]{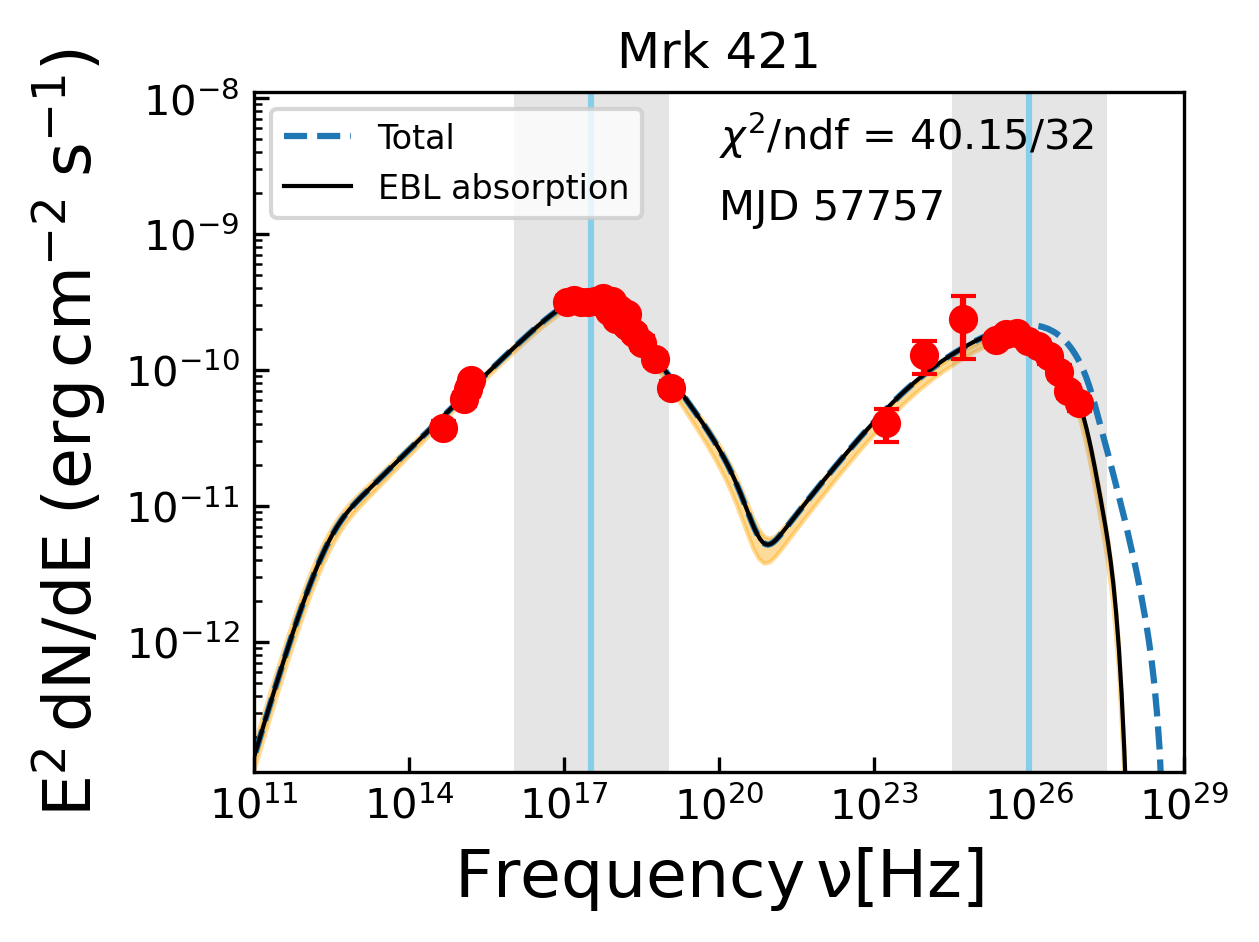}

    \includegraphics[width=0.244\textwidth]{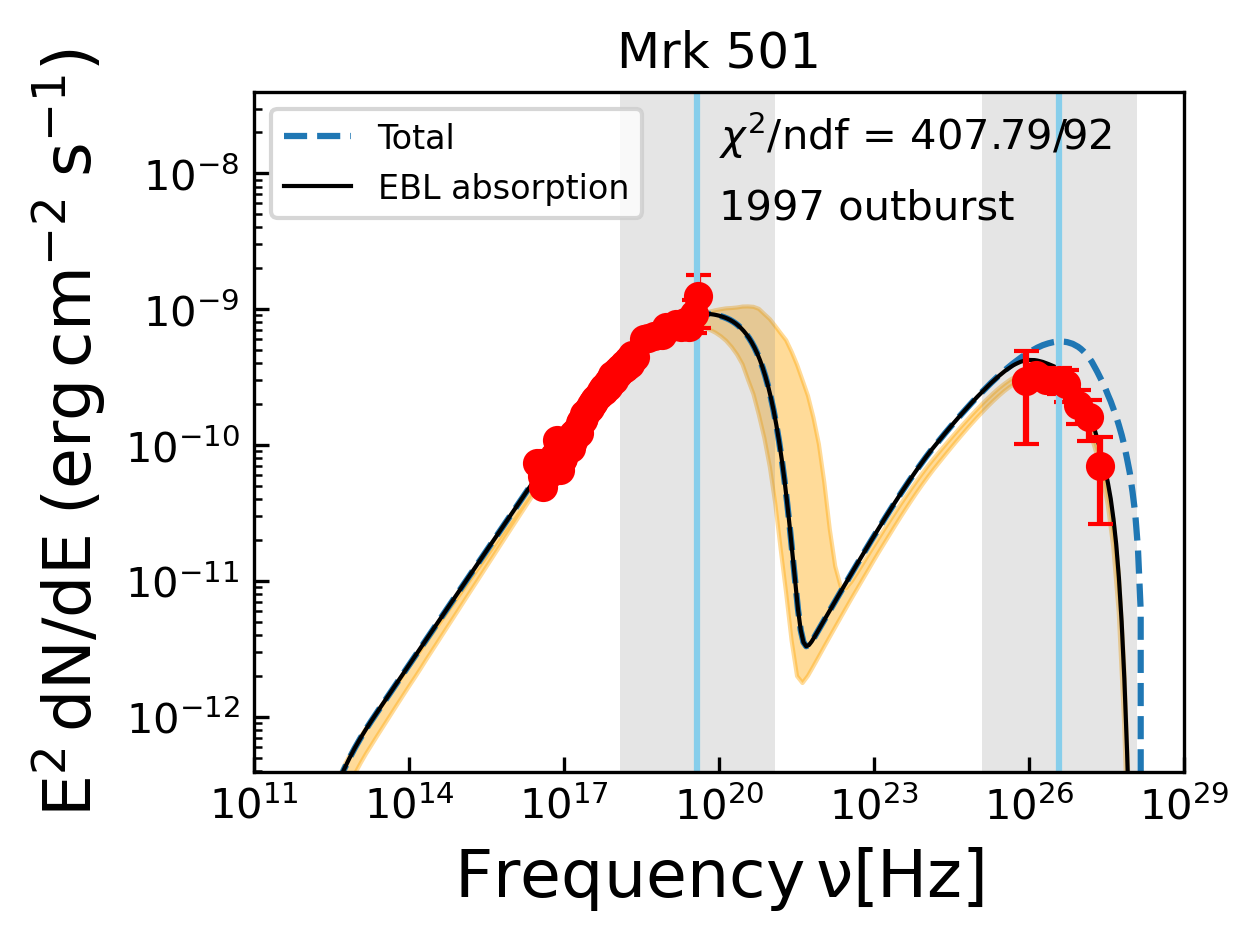}
    \includegraphics[width=0.244\textwidth]{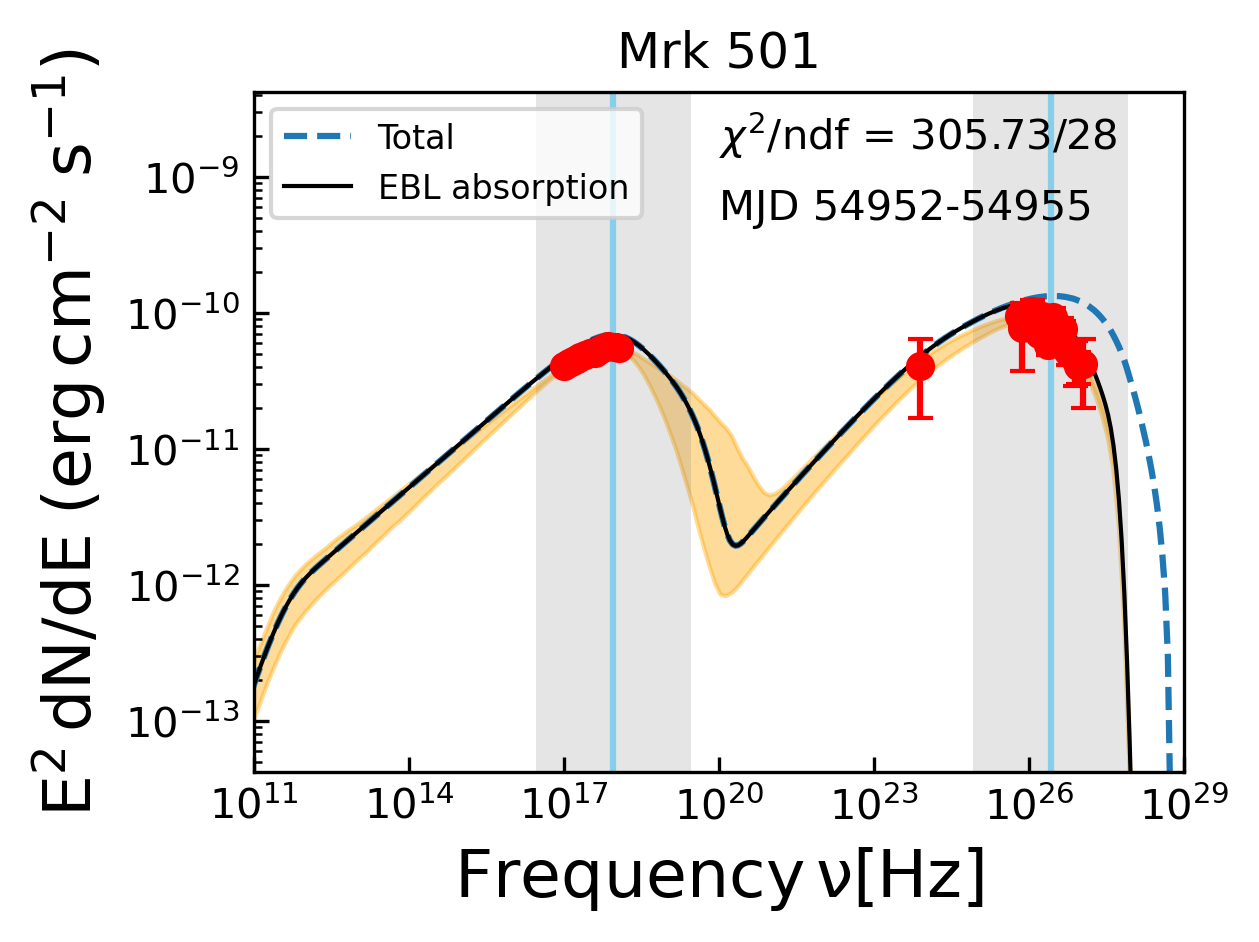}
    \includegraphics[width=0.244\textwidth]{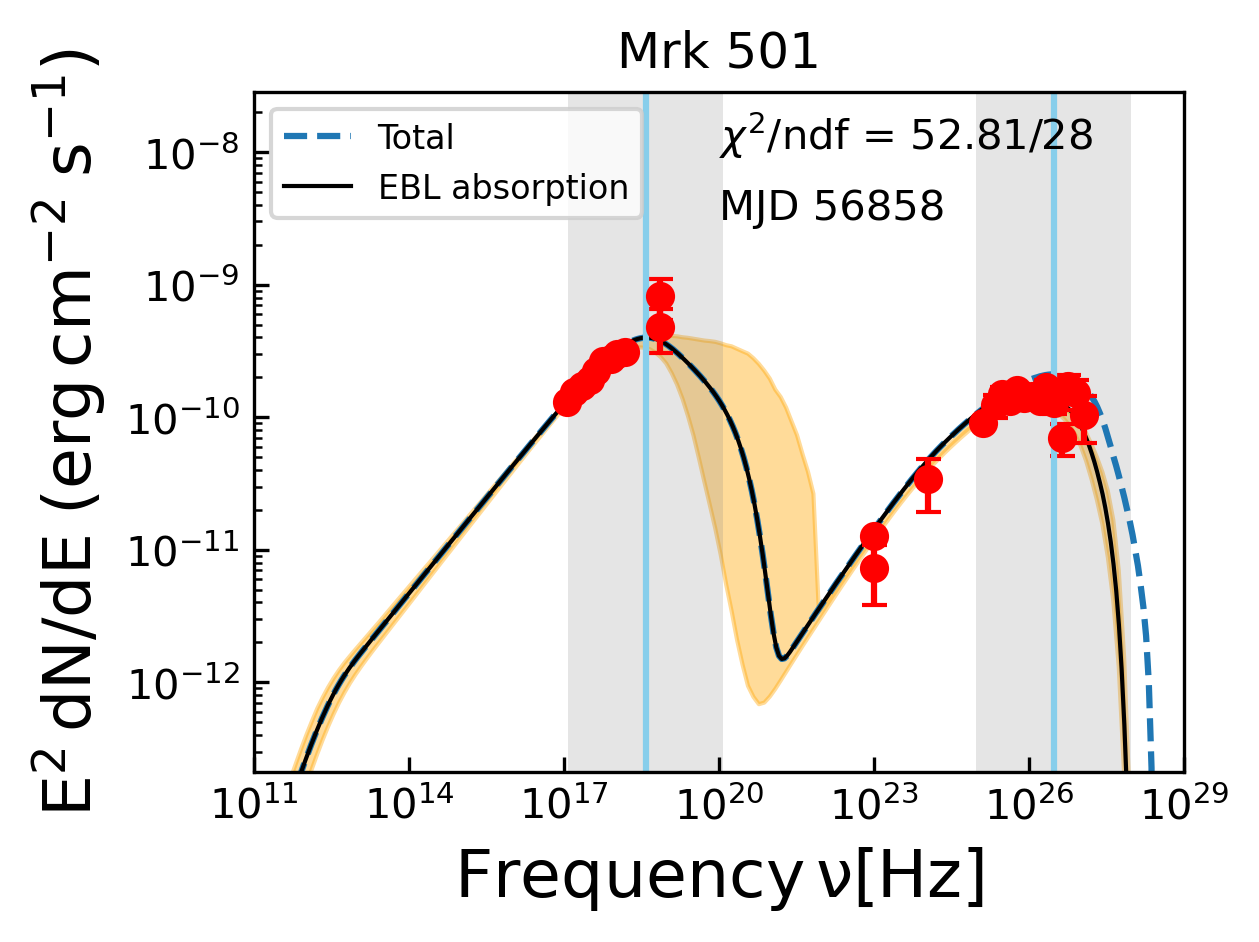}
    \includegraphics[width=0.244\textwidth]{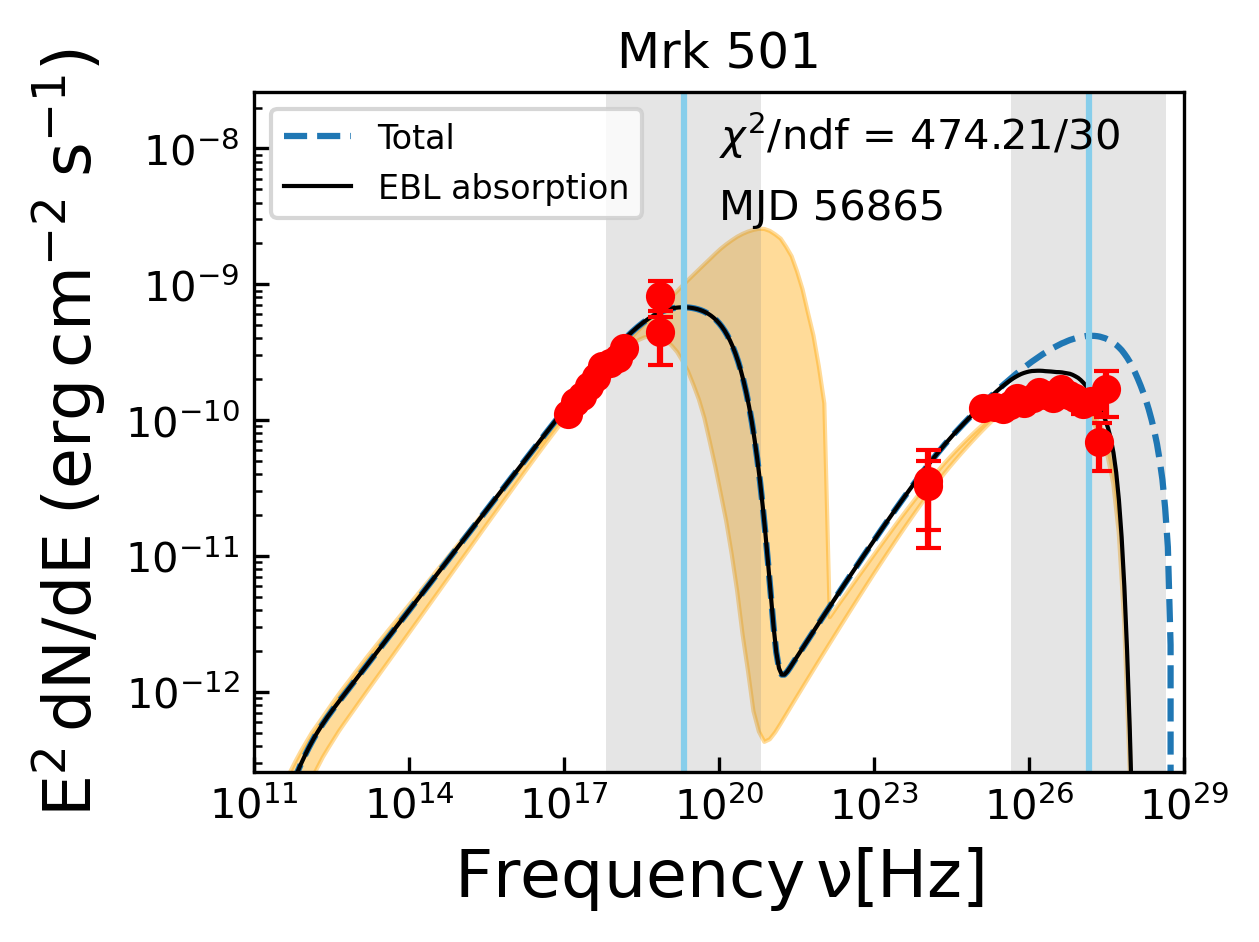}

    \includegraphics[width=0.244\textwidth]{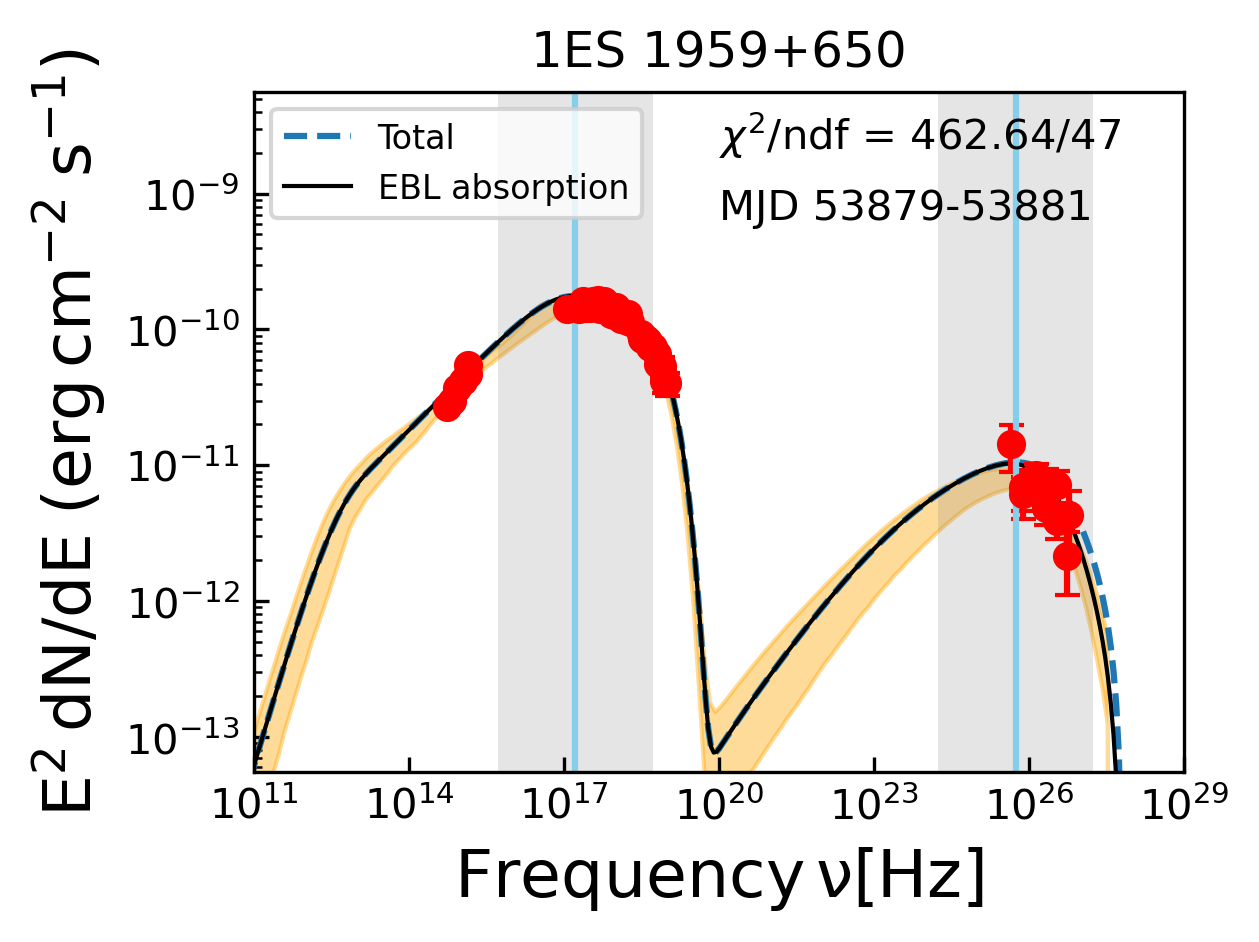}
    \includegraphics[width=0.244\textwidth]{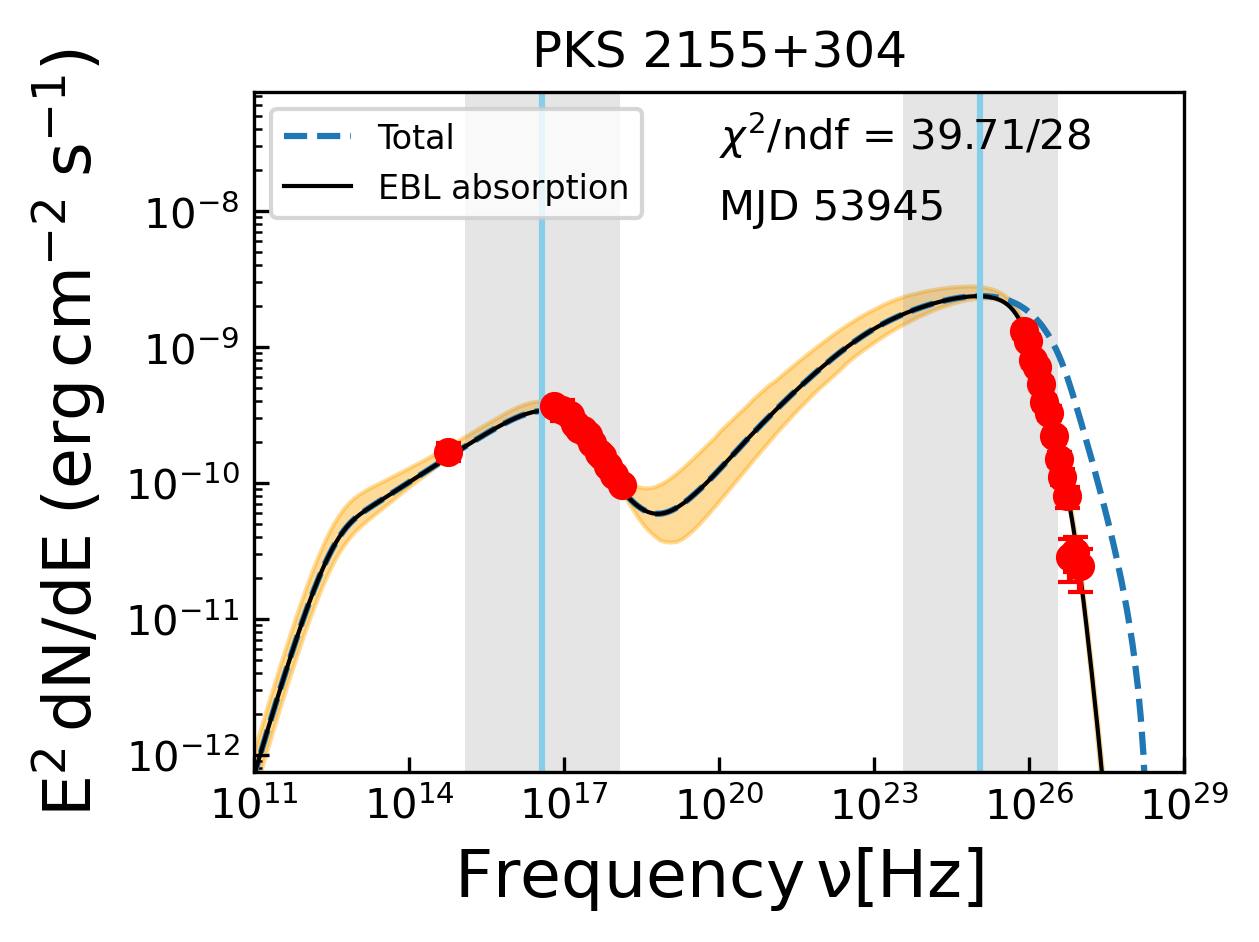}
    \includegraphics[width=0.244\textwidth]{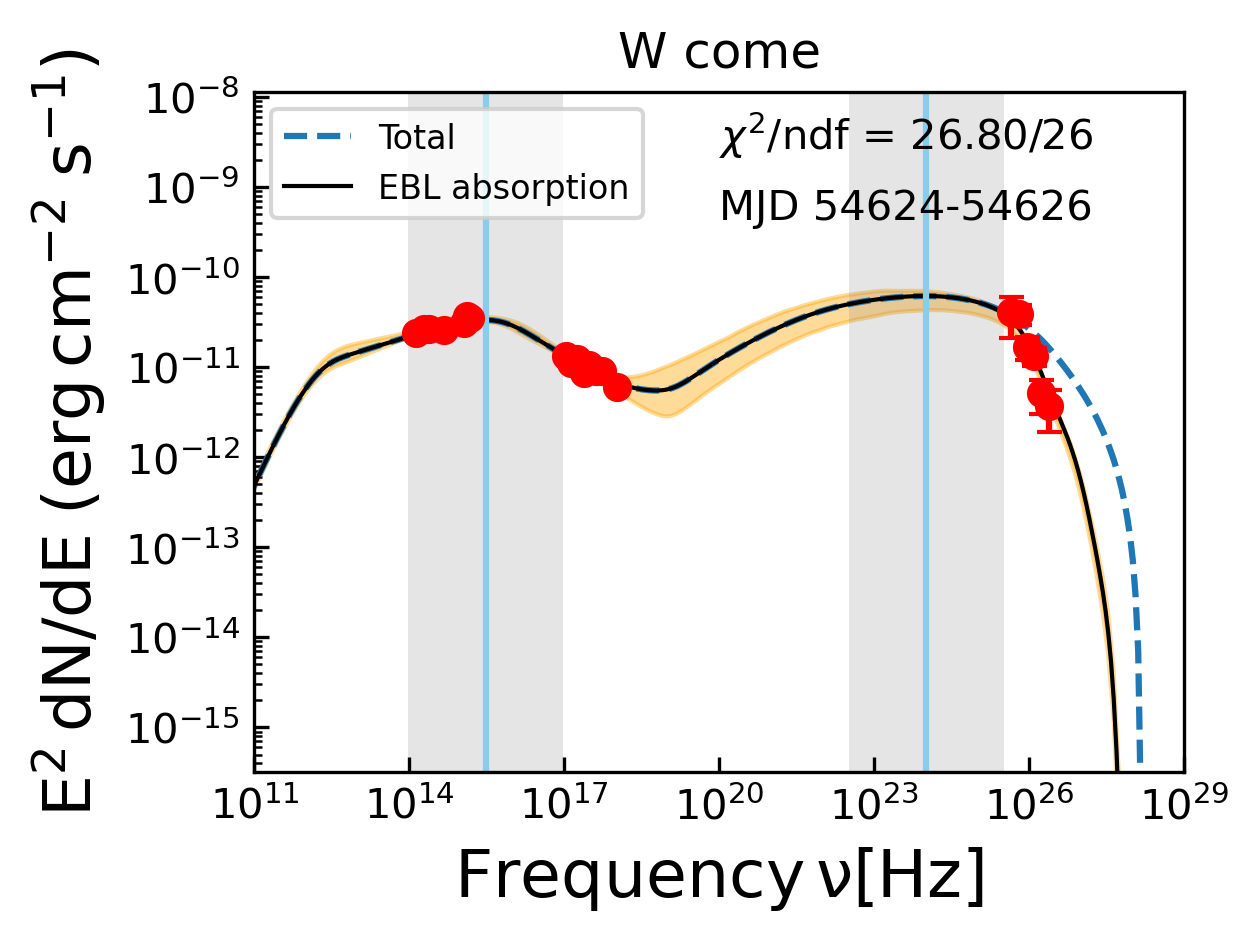}
    \includegraphics[width=0.244\textwidth]{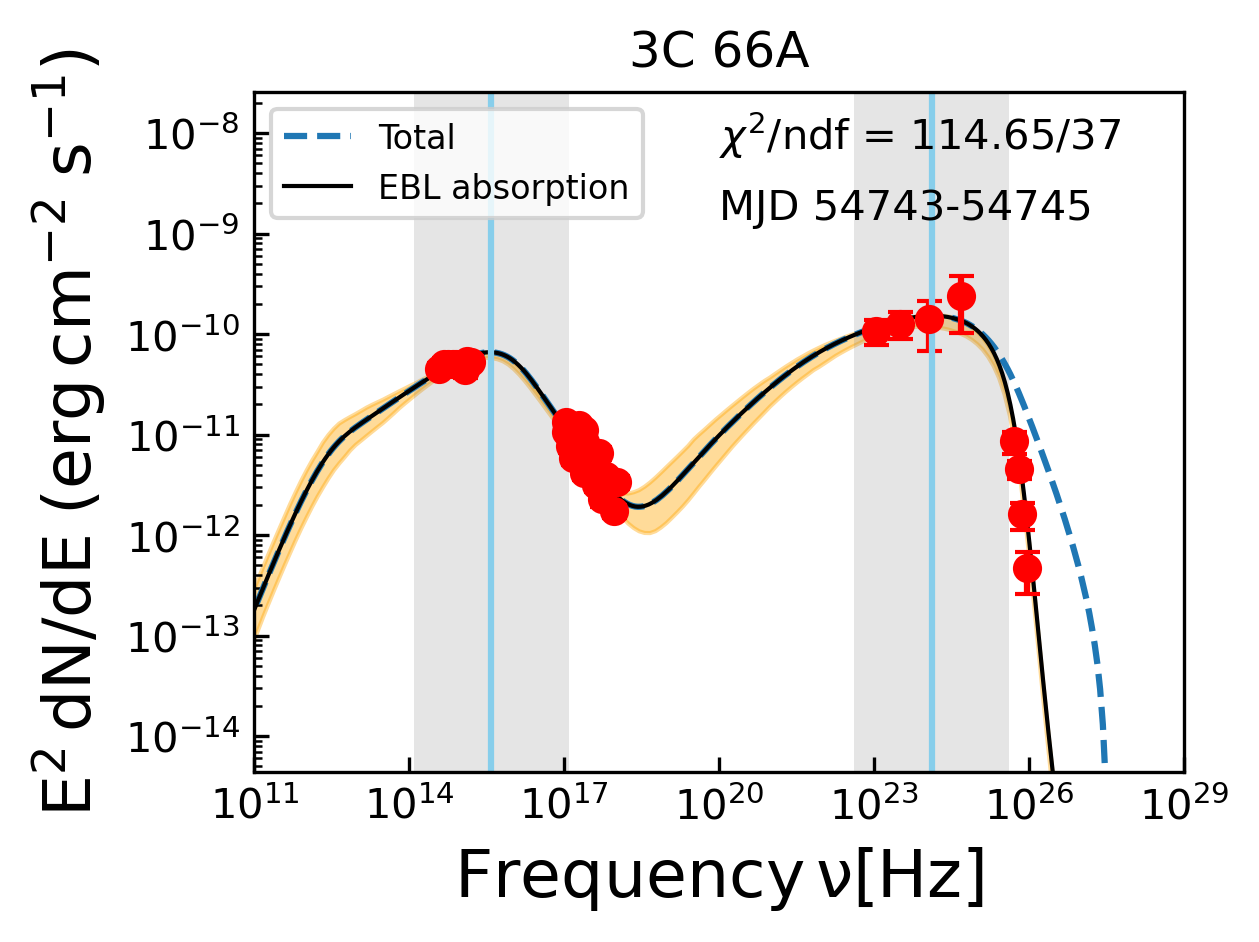}

    \includegraphics[width=0.244\textwidth]{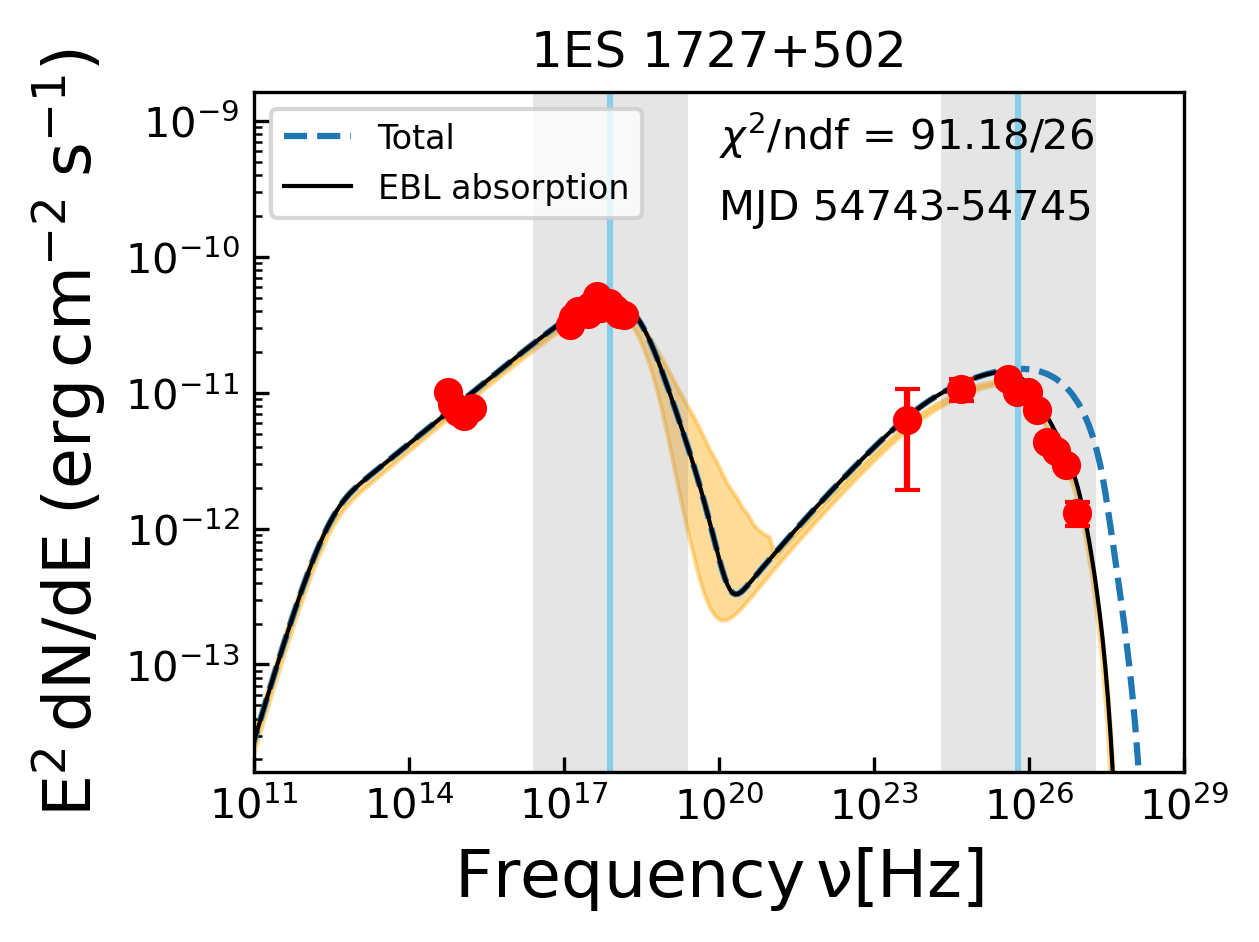}
    \includegraphics[width=0.244\textwidth]{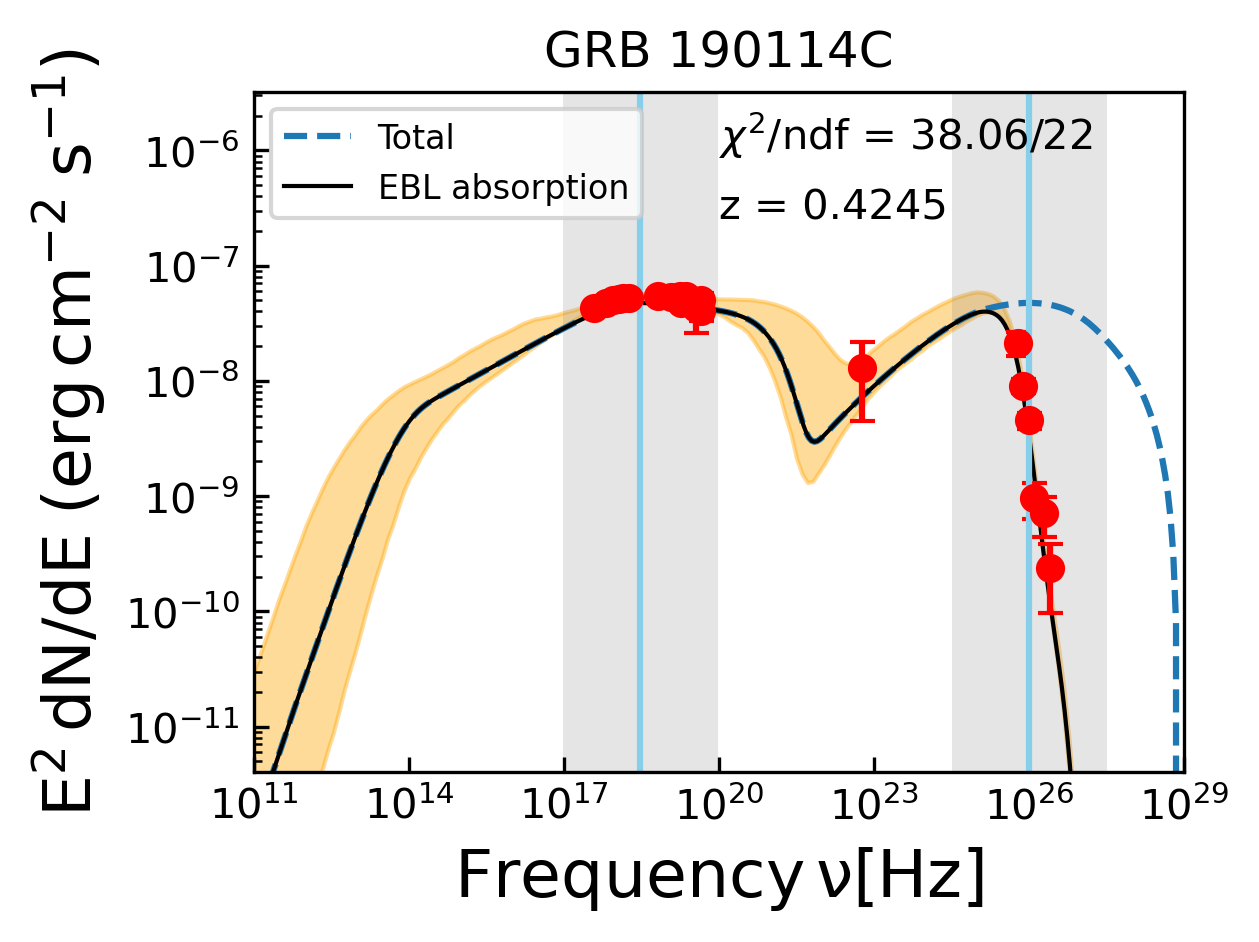}              \includegraphics[width=0.244\textwidth]{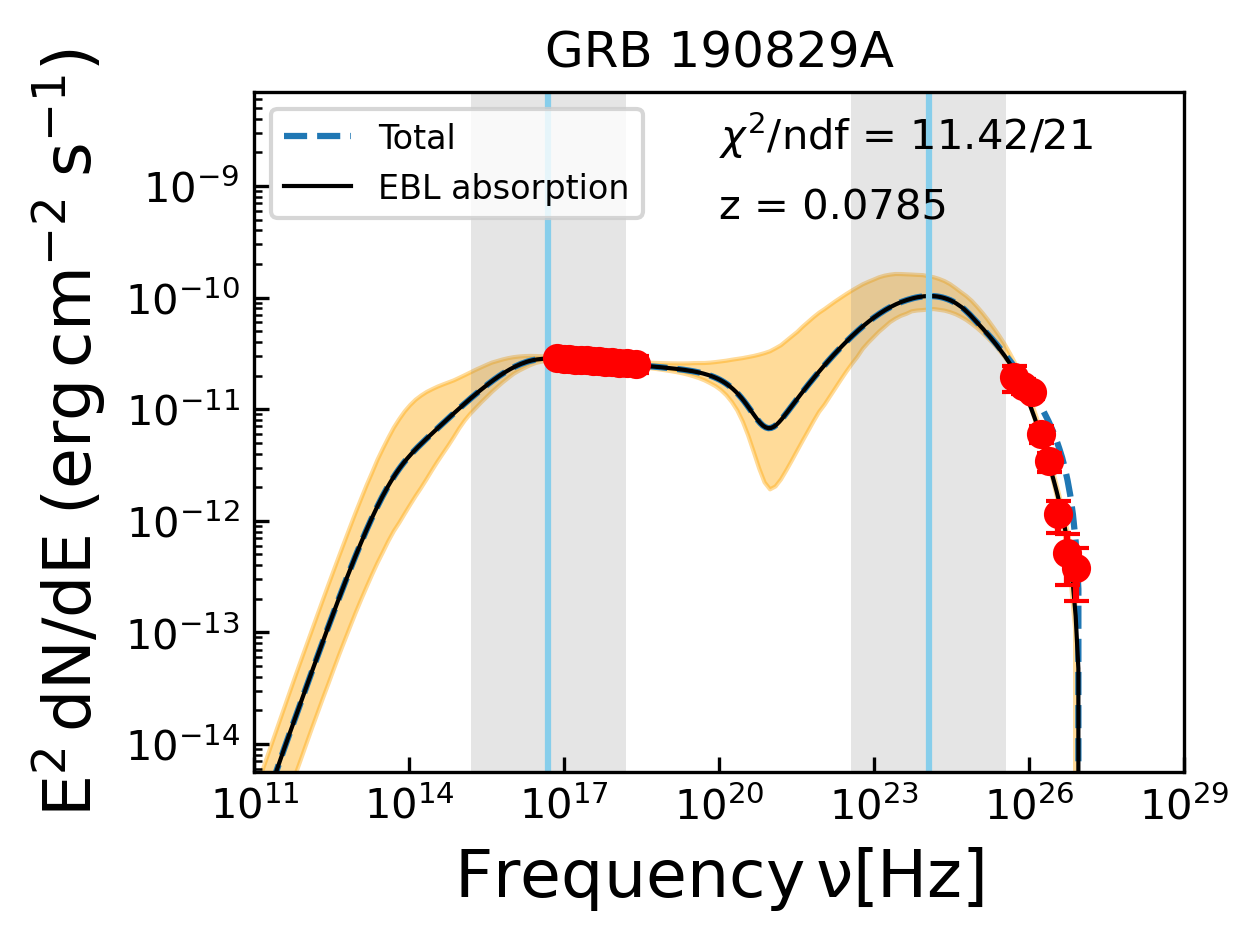}
    \includegraphics[width=0.244\textwidth]{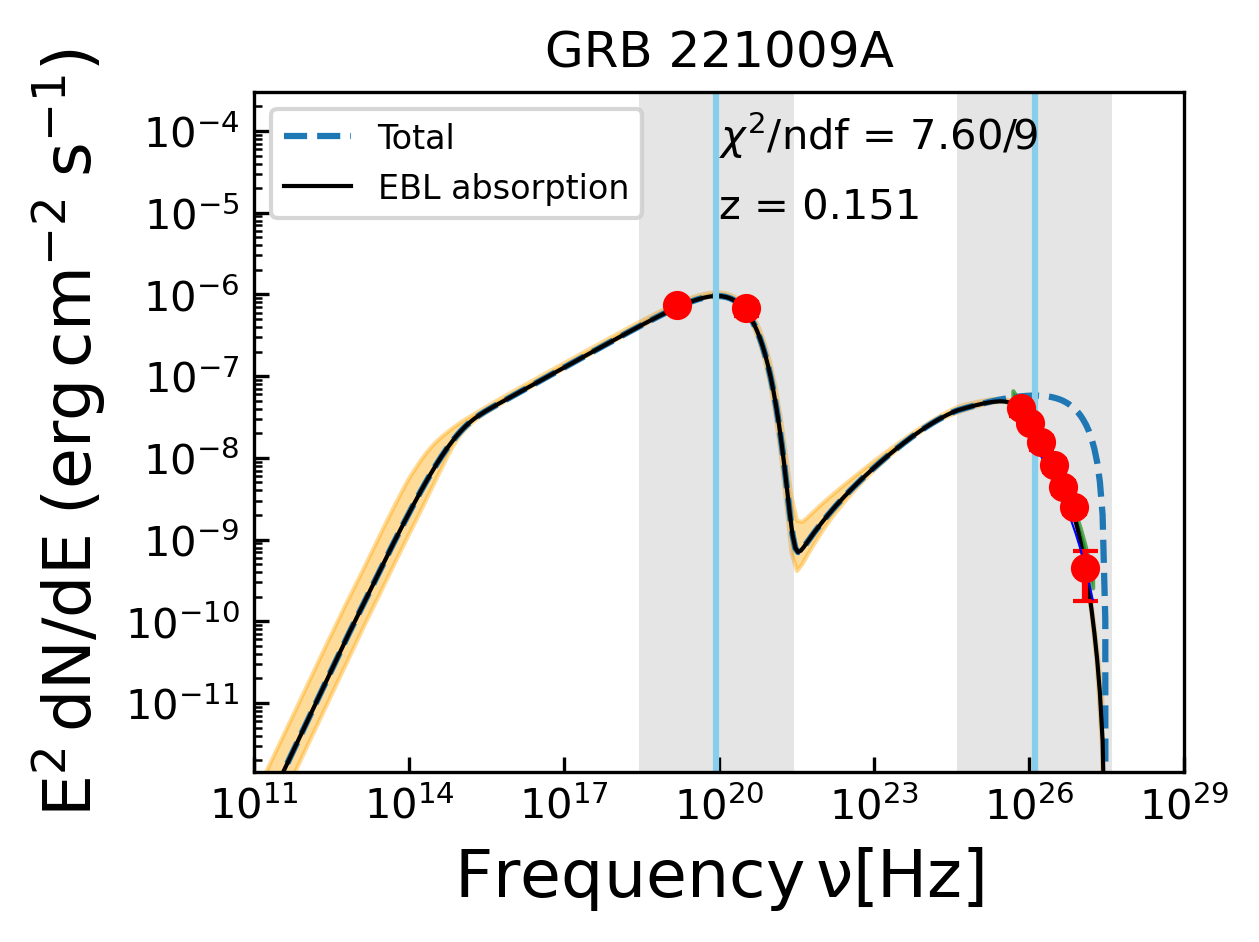}
    \caption{Comparisons of the best-fitting SEDs with the one-zone SSC model (solid line) and observed data (red points) of each source sample. Data are taken from the literature listed in the last column of Table \ref{tab:candi}. The confidence interval for the fitting is obtained with the MCMC fitting process.}
    \label{fig:sed}
\end{figure*}


\begin{table*}[h]
\tiny
    \centering
    \caption{Best-Fit Parameters of SED model}
\vspace{0.2cm}
    \begin{tabular}{lccccccccc}
        \hline
        Stat  & $\rm log_{10}B^{\prime} [G]$  & $\rm log_{10}\delta$ & $\rm log_{10}\gamma_{min}$  & $\rm log_{10}\gamma_{break}$ & $\rm log_{10}\gamma_{max}$ & $\rm p_1$ & $\rm p_2$ & $\rm log_{10}K_e$ & eletype \\
        \hline
		Mrk 421 \\
		55242-55245 & -0.67$\pm$0.21  & 0.80$\pm$0.09  & 2.78(fixed)  & 5.52$\pm$0.08  & 7.00(fixed)  & 1.50$\pm$0.09  & 4.11$\pm$0.10  & 42.50$\pm$0.16  & BPL\\
		55265 & -1.79$\pm$0.11  & 1.93$\pm$0.04  & 2.78(fixed)  & 5.66$\pm$0.06  & 6.60(fixed)  & 2.27$\pm$0.03  & 4.35$\pm$0.45  & 39.63$\pm$0.14  & BPL\\
		56130-56187 & -1.07$\pm$0.36  & 1.77$\pm$0.27  & 2.78(fixed)  & 4.92$\pm$0.26  & 6.20$\pm$0.57  & 2.23$\pm$0.12  & 5.18$\pm$0.40  & 41.23$\pm$0.70  & BPL\\
		57757 & -1.61$\pm$0.12  & 1.76$\pm$0.04  & 2.78(fixed)  & 5.45$\pm$0.07  & 6.80(fixed)  & 2.25$\pm$0.05  & 4.05$\pm$0.14  & 40.24$\pm$0.16  & BPL\\
		\hline
		Mrk 501 \\
		1997 outburst & -0.84$\pm$0.14  & 1.27$\pm$0.07  & 2.78(fixed)  & 5.93$\pm$0.22  & 6.86$\pm$0.31  & 1.88$\pm$0.06  & 2.93$\pm$0.91  & 39.48$\pm$0.55  & BPL\\
		54952-54955 & -2.52$\pm$0.38  & 1.83$\pm$0.09  & 2.78(fixed)  & 6.04$\pm$0.16  & 6.84$\pm$0.41  & 2.34$\pm$0.10  & 3.73$\pm$0.79  & 39.26$\pm$0.47  & BPL\\
		56858 & -1.36$\pm$0.32  & 1.53$\pm$0.11  & 2.78(fixed)  & 5.89$\pm$0.24  & 6.79$\pm$0.50  & 2.04$\pm$0.05  & 3.72$\pm$1.43  & 39.35$\pm$0.42  & BPL\\
		56865 & -2.02$\pm$0.68  & 1.70$\pm$0.19  & 2.78(fixed)  & 6.24$\pm$0.40  & 6.99$\pm$0.34  & 2.01$\pm$0.08  & 2.86$\pm$1.38  & 39.02$\pm$0.53  & BPL\\
		\hline
		1ES 1959+650 \\
		53879-53881 & -1.42$\pm$0.37  & 1.77$\pm$0.13  & 2.78(fixed)  & 5.00$\pm$0.34  & 5.96$\pm$0.15  & 2.23$\pm$0.34  & 3.20$\pm$0.23  & 40.28$\pm$0.94  & BPL\\
		\hline
		PKS 2155+304 \\
		53945 & -1.58$\pm$0.11  & 1.86$\pm$0.05  & 2.78(fixed)  & 5.00$\pm$0.10  & 6.44$\pm$0.60  & 2.51$\pm$0.12  & 4.16$\pm$0.21  & 42.22$\pm$0.31  & BPL\\
		\hline
		W comae \\
		54624-54626 & -2.06$\pm$0.23  & 1.96$\pm$0.09  & 2.78(fixed)  & 4.60$\pm$0.12  & 6.16$\pm$0.43  & 2.64$\pm$0.12  & 3.71$\pm$0.15  & 42.80$\pm$0.43  & BPL\\
		\hline
		3C 66A \\
		54743-54745 & -1.27$\pm$0.16  & 1.51$\pm$0.07  & 2.78(fixed)  & 4.53$\pm$0.24  & 6.11$\pm$0.67  & 2.29$\pm$0.24  & 4.49$\pm$0.15  & 45.03$\pm$0.31  & BPL\\
		\hline
		1ES 1727+502 \\
		57090–57440 & -1.56$\pm$0.10  & 1.75$\pm$0.04  & 2.78(fixed)  & 5.70$\pm$0.09  & 6.44$\pm$0.65  & 2.38$\pm$0.03  & 5.17$\pm$1.24  & 39.22$\pm$0.26  & BPL\\
		\hline
		GRB \\
		GRB 190114C & -1.28$\pm$1.26  & 1.96$\pm$0.45  & 3.31$\pm$0.47  & 5.50$\pm$0.47  & 7.00(fixed)  & 2.51$\pm$0.29  & 3.08$\pm$0.17  & 43.01$\pm$0.96  & BPL\\
		GRB 190829A & 0.44$\pm$0.37  & 0.46$\pm$0.12  & 3.00(fixed)  & 4.39$\pm$0.28  & 6.50(fixed)  & 2.11$\pm$0.64  & 3.07$\pm$0.04  & 44.07$\pm$0.74  & BPL\\
		GRB 221009A & 0.43$\pm$0.09  & 1.38$\pm$0.05  & 3.22$\pm$0.21  & -  & 6.07$\pm$0.03  & 2.30(fixed)  & -  & 54.68$\pm$0.07  & PL\\
		\hline
    \end{tabular}
    \label{tab:para}
\end{table*}

\begin{table*}[h]
\footnotesize
    \centering
    \caption{Fitting results}
\vspace{0.2cm}
    \begin{tabular}{lccccccc}
        \hline
        Stat   & $\rm log_{10}E_{peak}^{syn} [eV]$ & $\rm log_{10}E_{peak}^{ssc} [eV]$ & $\rm log_{10}L_{syn} [erg~s^{-1}]$  & $\rm log_{10}L_{ssc} [erg~s^{-1}]$   \\
        \hline
		Mrk 421 \\
		55242-55245 & $3.43\pm{0.01}$  & $11.45\pm{0.11}$  & $46.06\pm{0.01}$  & $45.99\pm{0.06}$ \\
		55265 & $3.54\pm{0.01}$  & $11.87\pm{0.11}$  & $45.99\pm{0.02}$  & $45.38\pm{0.02}$ \\
		56130-56187 & $2.48\pm{0.42}$  & $10.92\pm{0.11}$  & $46.03\pm{0.35}$  & $45.56\pm{0.03}$ \\
		57757 & $3.11\pm{0.01}$  & $11.56\pm{0.01}$  & $45.62\pm{0.01}$  & $45.42\pm{0.02}$ \\
		\hline
		Mrk 501 \\
		1997 outburst & $5.01\pm{0.32}$  & $12.19\pm{0.01}$  & $46.05\pm{0.08}$  & $45.73\pm{0.02}$ \\
		54952-54955 & $3.54\pm{0.01}$  & $12.08\pm{0.11}$  & $44.85\pm{0.03}$  & $45.18\pm{0.02}$ \\
		56858 & $4.17\pm{0.11}$  & $12.08\pm{0.11}$  & $45.63\pm{0.08}$  & $45.36\pm{0.03}$ \\
		56865 & $4.49\pm{0.26}$  & $12.82\pm{0.11}$  & $45.74\pm{0.13}$  & $45.55\pm{0.03}$ \\
		\hline
		1ES 1959+650 \\
		53879-53881 & $3.01\pm{0.26}$  & $11.35\pm{0.21}$  & $44.87\pm{0.03}$  & $43.58\pm{0.07}$ \\
		\hline
		PKS 2155+304 \\
		53945 & $2.27\pm{0.05}$  & $10.71\pm{0.26}$  & $46.81\pm{0.03}$  & $47.71\pm{0.06}$ \\
		\hline
		W comae \\
		54624-54626 & $1.11\pm{0.16}$  & $9.55\pm{0.32}$  & $45.70\pm{0.02}$  & $45.96\pm{0.11}$ \\
		\hline
		3C 66A \\
		54743-54745 & $1.11\pm{0.01}$  & $9.76\pm{0.16}$  & $47.31\pm{0.02}$  & $47.73\pm{0.04}$ \\
		\hline
		1ES 1727+502 \\
		57090–57440 & $3.43\pm{0.01}$  & $11.45\pm{0.01}$  & $45.35\pm{0.02}$  & $44.91\pm{0.01}$ \\
		\hline
		GRB \\
		GRB 190114C & $4.27\pm{0.21}$  & $11.56\pm{0.63}$  & $50.33\pm{0.01}$  & $50.35\pm{0.04}$ \\
		GRB 190829A & $2.27\pm{0.37}$  & $9.55\pm{0.58}$  & $45.41\pm{0.02}$  & $45.96\pm{0.16}$ \\
		GRB 221009A & $5.54\pm{0.01}$  & $11.77\pm{0.05}$  & $50.40\pm{0.02}$  & $49.26\pm{0.02}$ \\
		\hline
    \end{tabular}
    \label{tab:fit}
\end{table*}

Employing the model and method described above, we model fit simultaneous multiwavelength observations of data sample listed in the Table \ref{tab:candi}, and use MCMC to obtain the best fit results. Represented by Mrk 421, Figure \ref{fig:mcmc} presents the MCMC results of the best-fit parameters for the observed spectrum of a flaring event that occurred at \textit{MJD}=55275. The best SED model fitting parameters for each candidate are listed in Table \ref{tab:para}.

Figure \ref{fig:sed} illustrates the SED modeling fitting results, the first and second rows consist of four spectra each, representing the flaring states of Mrk 421 and Mrk 501 at four different times, followed by single-flare spectra of five other blazars, and finally, the afterglow radiation spectra of three GRBs. The solid red points represent the experimental data, with data points sourced from the literature listed in the second-to-last column of Table \ref{tab:candi} and references within. The blue dashed line represents the intrinsic energy spectrum before EBL absorption. The solid black line represents the best-fit spectra obtained under the constraints of these experimental data points. A clear bimodal structure can be observed in each plot, with the first peak corresponding to synchrotron radiation and the second peak generated by the SSC process. To quantitatively evaluate the fit quality, the goodness of fit and the number of free parameters for each fit are provided in each plot of Figure \ref{fig:sed}, demonstrating that all the SEDs can be well modeled by this simple leptonic SSC mechanism.


\begin{figure*}[!htp]
    \centering
    \includegraphics[width=0.49\textwidth]{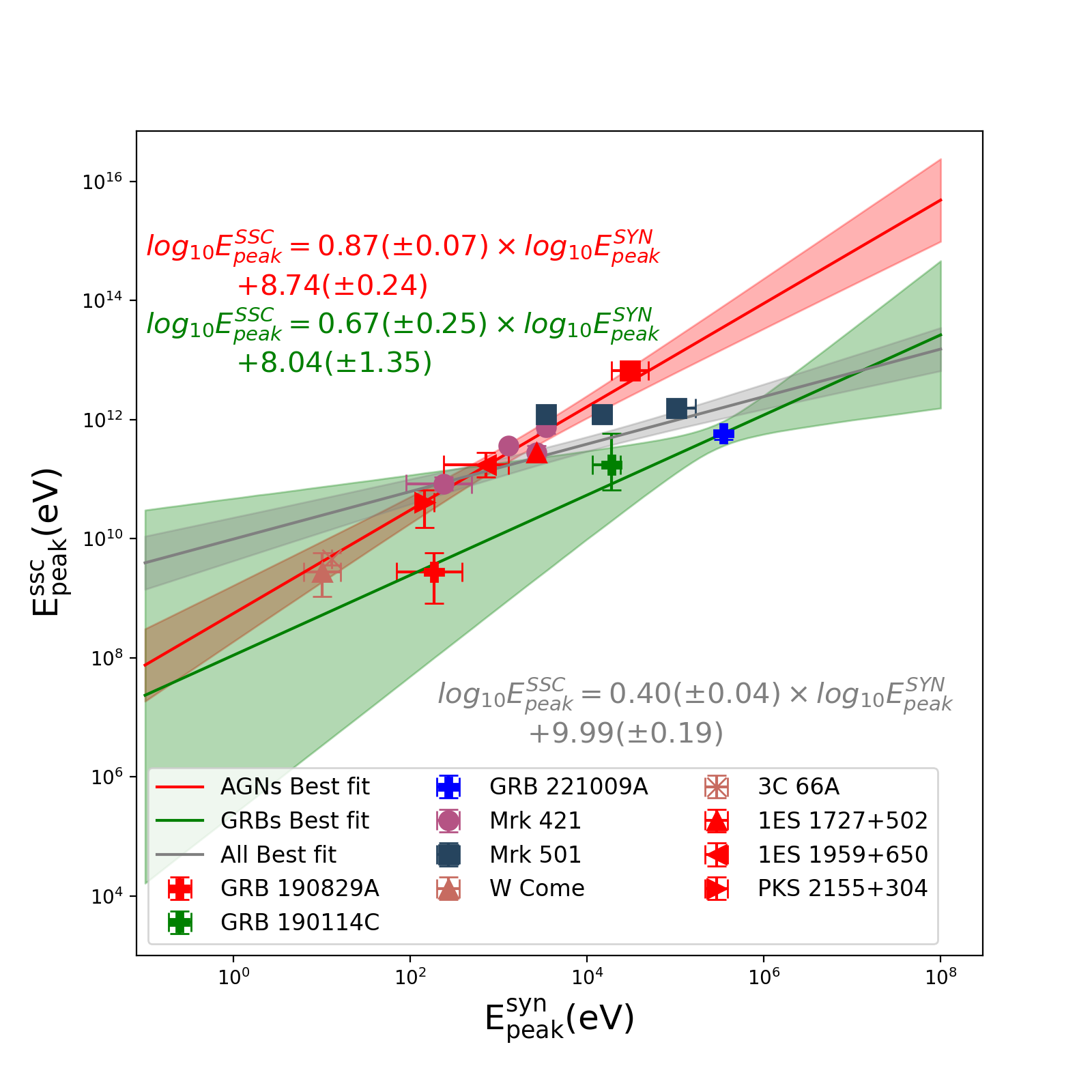}
    \includegraphics[width=0.49\textwidth]{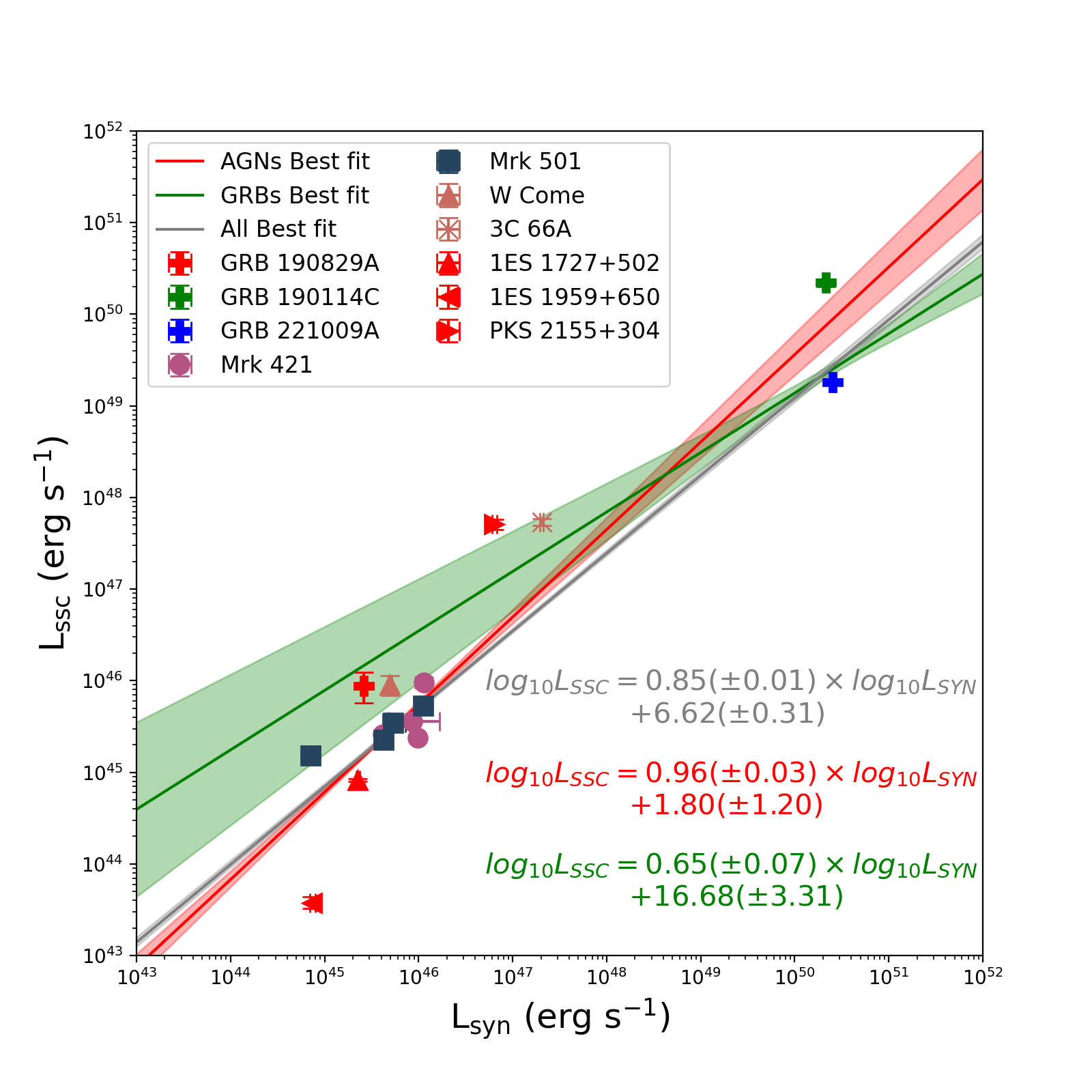}
    \caption{The correlations between $E_{peak}^{syn}$ and $E_{peak}^{ssc}$, $L_{syn}$ and $L_{ssc}$ for blazars and GRBs. The red and blue solid lines represent the best linear fits for blazars and GRBs separately, while the grey lines depict the combined fit results for both. The shaded regions indicate the 95\% confidence intervals of the spectral fitting. }
    \label{fig:rela}
\end{figure*}

The electrons responsible for SSC and synchrotron radiation are the same group of electrons, so it is expected that there would be a certain linear relationship between the peak position and peak height of the synchrotron and SSC processes, and the corresponding values for each source are also listed in Table \ref{tab:fit} and shown in the distribution plot in Figure \ref{fig:rela}.

The left plot of Figure \ref{fig:rela} displays the peak of two emissions of GRB and blazar. Fitting the peaks of synchrotron radiation and SSC with straight lines, $log_{10}E_{ssc}^{peak} = 0.87 \times log_{10}E_{syn}^{peak}+8.74$ and $ log_{10}E_{ssc}^{peak}=0.67 \times log_{10}E_{syn}^{peak}+8.04$ are obtained for blazar and GRBs, respectively. Although the energy of the accelerated particles inside the jet may be different, the peaks of SSC and synchrotron radiation both exhibit a good linear relationship. Additionally, the combined fit result for both blazars and GRBs is illustrated by the grey line, given by the fitting $log_{10}L_{ssc}=0.39\times log_{10}L_{syn}+10.01$.

The right plot of Figure \ref{fig:rela} illustrates the distribution of luminosity released by synchrotron and SSC radiation with equation \ref{eq:L}. A very clear linear relationship can be observed, which can be directly fitted with $ log_{10}L_{ssc}=0.95\times log_{10}L_{syn}+1.85$, $ log_{10}L_{ssc}=0.62\times log_{10}L_{syn}+18.31$ and $ log_{10}L_{ssc}=0.84\times log_{10}L_{syn}+6.95$ for blazars, GRBs, and both of them, respectively. It is evident that the emissions from blazars and GRB afterglows share energy budget similarities between these two radiation mechanisms. In other words, during the flaring state, the VHE blazars are primarily influenced by the scattering of the same group of electrons, and the TeV afterglow emissions could also be explained by the synchrotron + SSC mechanism.

\begin{figure*}[!htp]
    \centering
    \includegraphics[width=0.49\textwidth]{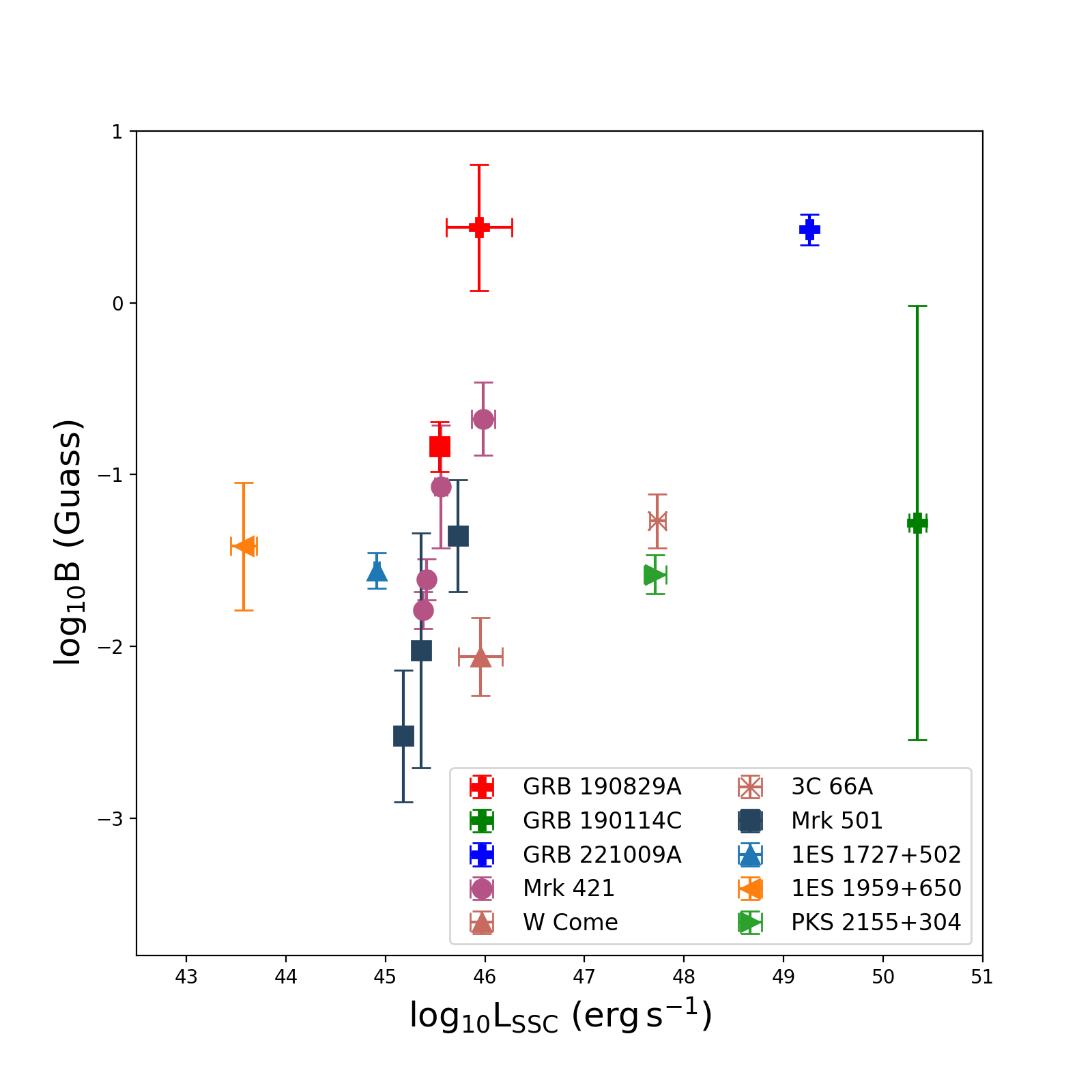}
    \caption{The correlations between magnetic filed and $\rm L_{ssc}$.}
    \label{fig:b_lumi}
\end{figure*}

The magnetic field and luminosity of the sources are depicted in Figure \ref{fig:b_lumi}. It is apparent that three GRBs are positioned in the upper right corner, suggesting that their magnetic fields are indeed larger than those of blazars. This is further supported by the peak of synchrotron emission from GRBs being higher than that from blazars, indicating the presence of a stronger magnetic field during the burst and potentially differing acceleration characteristics between the two.

\section{Summary}\label{sec:summary}

IACT observations of GRB 190114C and 180720B marked the beginning of VHE observations of GRBs, with LHAASO's observation of GRB 221009A further expanding into the 10 TeV observation window and, for the first time, tracing the entire emission from the onset of the burst. However, there remains a scarcity of GRB samples for studying their high-energy properties. GRBs and blazars are two populations of sources with many similar physical characteristics. This study explores the synchrotron and SSC emission properties of GRBs and blazars using a sample comprising 7 blazars and 3 GRBs with X-ray and VHE observations. By examining the correlations between $L_{syn}$ and $ L_{ssc}$, our findings reveal a robust linear relationship in the $ L_{syn}-L_{ssc}$ correlation. The similarities between GRBs and blazars, along with the connections in their synchrotron and SSC emissions, offer a pathway to investigate the highest energy range properties of GRBs. This approach makes use of the larger dataset from blazar observations and leverages the observational characteristics of low-energy synchrotron radiation. Additionally, we hope that LHAASO can observe more samples of GRBs to gather additional information for studying the properties of GRBs.

\acknowledgments
This work is supported by the National Natural Science Foundation of China  Nos. 12405124, 12275279, 12393851, 12393854, 12263007, the China Postdoctoral Science Foundation (No. 2023M730423).

\bibliographystyle{JHEP}
\bibliography{biblio.bib}
\end{document}